\documentclass[aps,pre,reprint, longbibliography, floatfix]{revtex4-1}
\usepackage{graphicx, mathrsfs, amsmath, amssymb, mathtools, amsfonts, bm, color, times, soul, cancel, physics}
\usepackage[colorlinks=true,pdfusetitle]{hyperref}
\usepackage[normalem]{ulem}
\definecolor{vivamagenta}{RGB}{190,52,85}
\hypersetup{allcolors=vivamagenta}

\makeatletter
\DeclareRobustCommand{\cev}[1]{{\mathpalette\do@cev{#1}}}
\newcommand{\do@cev}[2]{%
  \vbox{\offinterlineskip
    \sbox\z@{$\m@th#1 x$}%
    \ialign{##\cr
      \hidewidth\reflectbox{$\m@th#1\vec{}\mkern4mu$}\hidewidth\cr
      \noalign{\kern-\ht\z@}
      $\m@th#1#2$\cr
    }%
  }%
}
\makeatother

\newcommand{\s}{\mathsf{s}}
\newcommand{\ta}{\mathsf{t}}
\newcommand{\lowprime}[1]{ #1\raise0.4ex\hbox{$\scriptstyle\prime$} }
\newcommand{\lbra}[1]{\langle \hspace{-.7ex} \langle #1 \vert \hspace{-.45ex} \vert}
\newcommand{\lket}[1]{ \vert \hspace{-.45ex} \vert #1 \rangle \hspace{-.7ex} \rangle}

\begin{document}

\title{Uncovering nonequilibrium from unresolved events}
\author{Pedro E. Harunari}
\affiliation{Complex Systems and Statistical Mechanics, Department of Physics and Materials Science, University of Luxembourg, L-1511 Luxembourg, Luxembourg}

\date{\today}

\begin{abstract}
Closely related to the laws of thermodynamics, the detection and quantification of disequilibria are crucial in unraveling the complexities of nature, particularly those beneath observable layers. Theoretical developments in nonequilibrium thermodynamics employ coarse-graining methods to consider a diversity of partial information scenarios that mimic experimental limitations, allowing the inference of properties such as the entropy production rate. A ubiquitous but rather unexplored scenario involves observing events that can possibly arise from many transitions in the underlying Markov process--which we dub \textit{multifilar events}--as in the cases of exchanges measured at particle reservoirs, hidden Markov models, mixed chemical and mechanical transformations in biological function, composite systems, and more. We relax one of the main assumptions in a previously developed framework, based on first-passage problems, to assess the non-Markovian statistics of mutifilar events. By using the asymmetry of event distributions and their waiting-times, we put forward model-free tools to detect nonequilibrium behavior and estimate entropy production, while discussing their suitability for different classes of systems and regimes where they provide no new information, evidence of nonequilibrium, a lower bound for entropy production, or even its exact value. The results are illustrated in reference models through analytics and numerics.
\end{abstract}

\maketitle

%
%

\section{Introduction}\label{sec:introduction}

Built upon firm phenomenological roots, the celebrated theory of thermodynamics describes energy exchanges between systems, finding applications across a plethora of fields, from the scales of single particles to that of black holes. In contrast to statistical mechanics, it finds its merit in coarse-graining microscopic degrees of freedom, which leads to macroscopic descriptions that require the monitoring of promptly accessible quantities.

Developed in the past 30 years, stochastic thermodynamics applies a thermodynamic interpretation to Markov processes, energetically characterizing jumps between mesostates. While ``traditional'' thermodynamics works at thermal equilibrium, formalized by the detailed balance condition, the stochastic version distinguishably does not. A thermodynamic treatment of nonequilibrium systems is much needed to understand nature since living systems constantly dissipate energetical resources to generate order, and technological applications leverage them to output work. Detecting disequilibria is revealing; witnessing broken detailed balance becomes evidence for energy dissipation, which possibly points out the consumption of resources, production of waste, constant exchanges of matter, and might unfold in thermodynamic implications regarding control, optimization, presence of hidden pathways, adaptation, and more. It probes the intricacies of internal arrangements and force balances, and thus is the focal point of many theoretical and experimental developments~\cite{fangNonequilibriumPhysicsBiology2019, gnesottoBrokenDetailedBalance2018, zollerEukaryoticGeneRegulation2022, hartichNonequilibriumSensingIts2015, yangPhysicalBioenergeticsEnergy2021}.

Entropy production rate (EPR), a paramount quantity of nonequilibrium thermodynamics, does precisely the detection and quantification of distance to equilibrium, and is present in stochastic thermodynamics' own prominent results: fluctuation theorems \cite{PhysRevLett.78.2690, PhysRevE.60.2721, PhysRevLett.71.2401}, thermodynamic uncertainty relations \cite{PhysRevLett.114.158101, horowitz2020thermodynamic}, speed limits \cite{PhysRevLett.125.120604, PhysRevLett.121.070601}, nonequilibrium responses \cite{PhysRevX.10.011066, Basu_2015}, to name but a few. Beyond the violation of fluctuation-dissipation relations, measuring a nonzero EPR is clear evidence of nonequilibrium. Notably, the nonnegativity of the mean EPR generalizes the second law, lifting it to a status similar to that of entropy in traditional thermodynamics.

\begin{figure}[t]
    \centering
    \includegraphics[width=.7\columnwidth]{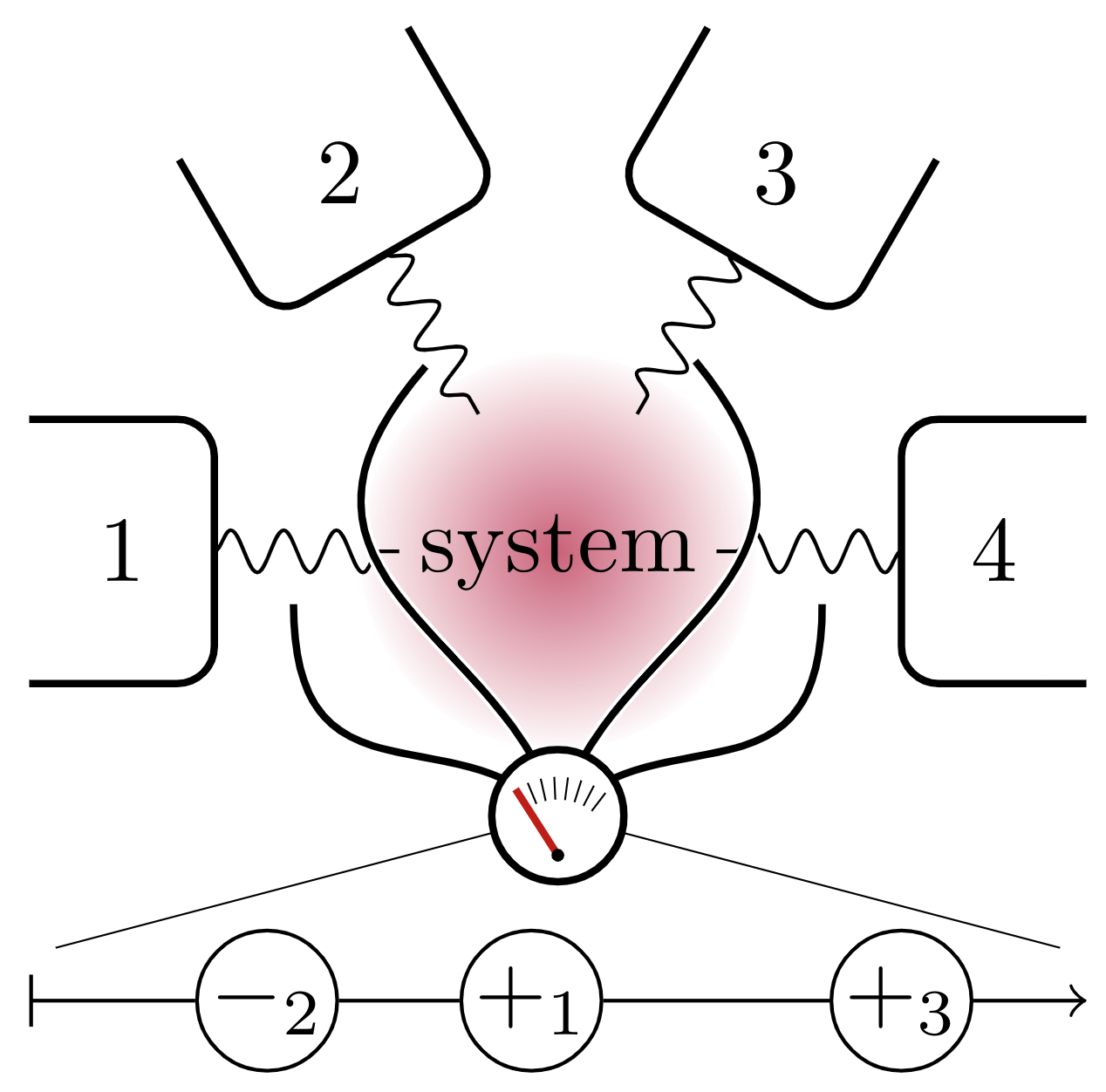}
    \caption{Exchanges between a system and reservoirs are monitored rather than the internal transitions of the system itself, yielding a trajectory that still carries thermodynamic information about the system. The squiggly arrows represent the monitored interactions that provide a trajectory comprised of multifilar events, represented as circles on a timeline.}
    \label{fig1}
\end{figure}

More akin to statistical mechanics than to traditional thermodynamics, stochastic thermodynamics struggles when the said mesoscopic states are coarse-grained; for instance, there is no obvious path to EPR when measurements are not performed at the level of individual edges. Resolving individual edges and requiring visibility of all of them is a major setback for the application of stochastic thermodynamics to systems beyond toy models or specifically tailored experiments. Very recently, different methods have been explored to infer its value in a variety of scenarios, usually providing lower bounds. Some methods involve: an integral fluctuation theorem for hidden EPR due to variable separation \cite{PhysRevE.88.022147, herpichEffectiveThermodynamicsTwo2020}, thermodynamic uncertainty relations for current precision \cite{PhysRevLett.124.120603, PhysRevE.100.060102, li2019quantifying, PhysRevE.101.042138}, numerically solving an optimization problem \cite{skinner2021improved} or establishing masked dynamics \cite{PhysRevE.91.012130, Bisker_2017, Ehrich_2021, blom2023milestoning}, which can involve quantifying asymmetries in forward and backward waiting-times distributions \cite{PhysRevLett.105.150607, martinez2019inferring}, time-scale separation \cite{PhysRevE.85.041125, boEntropyProductionStochastic2014, rahavFluctuationRelationsCoarsegraining2007}, the statistics of transitions when states are hidden \cite{PhysRevX.12.041026, PhysRevX.12.031025, PhysRevLett.130.257101, PhysRevE.107.L042105, garilli2023fluctuation, pietzonkaThermodynamicCostPrecision2023}, and others \cite{maesFreneticBoundsEntropy2017, berezhkovskii_forwardbackward_2019, cristadoroRecurrenceTimesWaiting2023, pagareStochasticDistinguishabilityMarkovian2024a, singh2023inferring, ferricortés2023entropy, liang_thermodynamic_2023, ertel2023estimator, degunther2023fluctuating, yuInversePowerLaw2021, baiesiEffectiveEstimationEntropy2023, hartichEmergentMemoryKinetic2021, lucenteRevealingNonequilibriumNature2023, gnesottoLearningNonequilibriumDynamics2020, lynnDecomposingLocalArrow2022, tanScaledependentIrreversibilityLiving2021}. See also a more detailed overview of recent approaches to nonequilibrium detection in Ref.~\cite{lucenteInferenceTimeIrreversibility2022}.

Although the dynamics of mesoscopic systems is typically observed through changes in the environment, ``from outside'', many of the mentioned methods fall short since a change in configuration is not necessarily associated with a unique change in the environment. As a paradigmatic example, consider a system in contact with many reservoirs driving it away from equilibrium through the exchange of physical quantities; some examples are transport problems \cite{PhysRevLett.120.087701, PhysRevB.104.195408, Cuetara_2015}, photon detections \cite{Viisanen_2015}, Maxwell demons \cite{PhysRevLett.123.216801, PhysRevE.103.032118} and chemical reaction networks \cite{rao2018conservation}. Also, consider that the experimenter does not directly monitor the system, since it might be beyond the experimental limitations or too sensitive to perturbations. See Fig.~\ref{fig1} for an illustration. The measurements open the possibility of inferring system properties, such as the distance to equilibrium, but they pose additional challenges. Several distinct transitions between system mesostates usually give rise to the same event in a reservoir, and thus we dub them multifilar (in contrast to the unifilar case in which an observable event is composed of a unique transition), rendering the trajectories non-Markovian. In this case, some of the previously known results do not apply, and the problem quickly becomes numerically expensive. Some other examples of systems whose observables have multifilar nature are molecular motors with distinct pathways causing mechanical movement \cite{bierbaumChemomechanicalCouplingMotor2011, abbondanzieriDirectObservationBasepair2005, chemlaExactSolutionsKinetic2008}, jumps between metastable states, the composite operation of identical indistinguishable subsystems, chemical reaction networks, and the observation of state occupancy in systems that can have more than one transition between the same pair of states.

In this contribution, we extend the framework for the statistics of a partial set of observable transitions \cite{PhysRevX.12.041026, PhysRevX.12.031025} to the more wide-ranging case of multifilar events and develop paths to detect and quantify nonequilibrium behavior in distinct scenarios using minimal assumptions. We start by formalizing the mathematical framework, obtaining analytical expressions for sequences of multifilar events and waiting-times based on first-passage problems, both in the semi-Markov approach and the fully non-Markovian case. We proceed by showing that the probabilities of each individual event compared to its time reverse bounds the EPR from below, this can be done with no prior knowledge of the system itself and can be applied to situations where the physical characteristics of the system and reservoirs are unknown or inaccessible. Second, we still do not require prior knowledge and consider a semi-Markov approach that uses conditional probabilities and waiting-time distributions to bound the EPR from below, which can be viewed as the generalization of the EPR estimation in \cite{PhysRevX.12.041026, PhysRevX.12.031025}. Stemming from the latter, we put forward quantities that can detect the presence of nonequilibrium behavior in sequences of multifilar events, including their waiting-times. These model-free estimators are particularly relevant to assess the nonequilibrium thermodynamics of partially observed systems since they do not include the restrictive assumption that each observable is generated by a single transition. Third, if some prior knowledge of the system model and topology is present, which can come in the form of e.g. possible chemical reactions or physical structure of a device, we put forward a simple method that measures the affinity and consequently provides the exact EPR, bypassing the need for establishing a model and quantifying its transition rates. We describe scenarios to decide the suitability of each estimator and the conditions that hinder their relevance. The results are illustrated in relevant models, and we conclude by discussing the results and open questions.

%
%

\section{Framework for multifilar events}\label{sec:2}

We consider a continuous-time Markov chain defined over an irreducible state space whose dynamics is captured by time-independent transition rates, establishing a master equation
\begin{equation}
    d_t \mathbf{p}(t) = \mathbf{R} \mathbf{p}(t),
\end{equation} 
here \(\mathbf{R}\) is the rate matrix with entries \(R_{ij}\) being the transition rate from state \(j\) to \(i\) if \(i\neq j\), or minus the exit rate \(R_{ii} = - \sum_{j} R_{ji}\) otherwise. A pair of states \(i \neq j\) can be connected by more than one transition, rendering the state space a multigraph, which can be expressed as
\begin{equation}
    R_{ij} = \sum_\ell r_\ell \delta_{i, \ta(\ell)} \delta_{j, \s(\ell)},
\end{equation}
with the sum spanning over the set \(\mathcal{L}\) of all possible transitions. In our notation, \(\ell \in \mathcal{L}\) is a transition from source state \(\s(\ell)\) to target state \(\ta(\ell)\), and its rate is rate \(r_\ell\). The probability distribution will reach a unique stationary value at long times that will be, for simplicity, referred to as \(\mathbf{p}\) since hereafter we assume stationarity.

In this work we consider that a subset of all transitions can be detected, namely the observable transitions \(\mathcal{L}_\text{obs} \subseteq \mathcal{L}\). We assume microreversibility of all possible transitions \(\mathcal{L}\), i.e. if \(\ell \in \mathcal{L}\) then \(\bar{\ell} \in \mathcal{L}\). This assumption can be relaxed by considering ``hidden irreducibility''~\cite{PhysRevE.107.L042105}, which ensures that any two observable transitions can occur consecutively, and preserving the thermodynamic consistency of all observables.

\begin{figure}
    \centering
    \includegraphics{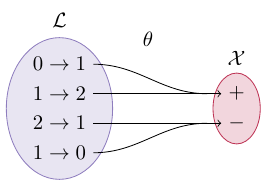}
    \caption{Example of multifilar events: Whenever a three-level system exchanges electrons with a metallic lead, a pair of transitions from space \(\mathcal{L}\) is involved with a single event observed on the lead in \(\mathcal{X}\). The map \(\theta\) relating both is non-injective.}
    \label{fig:contraction}
\end{figure}

A crucial step for this work is accounting for multifilar events: events that immediately emerge from the occurrence of more than one transition. As yet another example, consider a three-level system that exchanges matter with a lead; when an electron is transferred to the system, a jump can occur from the ground to the first excited state, or from the first to the second excited state, but the same increment is observed in the current system-lead in both cases. If energy exchanges are monitored, the multifilar character is present if the energy gaps between the three states are the same, otherwise it would be possible to distinguish the internal transitions. See Fig.~\ref{fig:contraction}.

The relation between observable transitions and their respective multifilar events is given by the map \(\theta : \mathcal{L}_\text{obs} \to \mathcal{X}\), supported by the subset of all observable transitions. This map is non-injective; otherwise, it would not establish multifilar events. We also assume that each transition is associated with at most one multifilar event; otherwise, \(\theta\) would not be a map. Since the reverse of an observable transition is also observable, the reverse of an element \(x \in \mathcal{X}\) is also a multifilar event \(\bar{x} \in \mathcal{X}\). As should, the coarse-graining map commutes with time reversal: \(\theta \bar{\ell} = \bar{x}\) for \(\theta \ell = x\). If the timeseries of observable transitions is
\begin{equation}
    \ell_0 \xrightarrow{t_1} \ell_1 \xrightarrow{t_2} \cdots  \xrightarrow{t_{n}} \ell_n,
\end{equation}
then the observation of multifilar events yields
\begin{equation}
    \theta\ell_0 \xrightarrow{t_1} \theta\ell_1 \xrightarrow{t_2} \cdots \xrightarrow{t_{n}} \theta\ell_n.
\end{equation}
In principle, a multifilar event is a label given to one or more observable transitions produced by the Markov process.

\subsection{Semi-Markov}

Even though the sequence of events is non-Markovian, a semi-Markov treatment is relevant for practical purposes. Empirical estimation of the statistics of more than two events often becomes computationally challenging or requires the acquisition of a lot of data. For this reason, we start exploring the statistics between a pair of events, which can be later used to probe collected data or to evaluate quantities under the semi-Markov assumption.

Immediately after a multifilar event \(x\), the probability distribution in the state space is a statistical mixture of all possible targets of transitions in the subset \(\theta^{-1} x \subseteq \mathcal{L}_\text{obs}\). The non-delta distribution after an event introduces non-Markovianity in the sequence of multifilar events. By comparing the rates at which transitions occur, this distribution can be expressed as
\begin{equation}
    \mathbf{p}_{\vert x} = \boldsymbol{\Gamma}_{x} \mathbf{p},
\end{equation}
where we define the event-propagating matrix
\begin{equation}\label{eq:Gammamat}
    \boldsymbol{\Gamma}_{x} \coloneqq \frac{1}{\mathcal{K}_x} \sum_{\ell \in \theta^{-1}x} r_\ell \boldsymbol{\Delta}_\ell.
\end{equation}
We also define \(\boldsymbol{\Delta}_\ell\) as a square matrix of size equals to the cardinality of the state space and whose elements are zeros but for 1 in the element of row \(\ta(\ell)\) and column \(\s(\ell)\) \footnote{Connecting to the notation of Ref.~\cite{PhysRevX.12.041026}, \(\boldsymbol{\Delta}_\ell\) can be expressed as \(\lket{\ell} \lbra{\ell}\).}, and \(\mathcal{K}_x \coloneqq \sum_{\ell \in \theta^{-1}x} r_\ell p_{\s(\ell)} \) as the traffic rate of event \(x\) (also known as its dynamical activity, a time-symmetric observable). The observable traffic rate, which measures the number of all events per unit time, is then \(\mathcal{K}_\text{obs} \coloneqq \sum_x \mathcal{K}_x \leq \mathcal{K}\), where \(\mathcal{K}\) is the total traffic rate that also includes the hidden transitions.

The state space propagation between two observable multifilar events can be obtained by solving a first-passage problem in an auxiliary dynamics \cite{PhysRevX.12.041026, sekimoto2021derivation}. Let \(\mathbf{S} \coloneqq \mathbf{R} - \sum_{\ell \in \mathcal{L}} r_\ell \boldsymbol{\Delta}_\ell\) be the survival propagator, then \(\exp (t \mathbf{S})\) propagates the probability distribution for a duration \(t\) conditioned on the absence of observable transitions and \(r_\ell [\exp (t \mathbf{S}) \mathbf{p}]_{\s(\ell)}\) is the probability of performing transition \(\ell\) after time \(t\) starting from any \(\mathbf{p}\). The survival matrix can also be expressed in terms of Eq.~\eqref{eq:Gammamat}:
\begin{equation}\label{eq:survivalmatrix}
    \mathbf{S} = \mathbf{R} - \sum_{x \in \mathcal{X}} \mathcal{K}_x \boldsymbol{\Gamma}_x.
\end{equation}

At the level of observable transitions, we can now obtain the probability of transition \(\ell\) occurring after time \(t\) has passed since the multifilar event \(x\),
\begin{equation}
    P(\ell, t \vert x) = r_\ell \left[ e^{t \mathbf{S}} \boldsymbol{\Gamma}_x \mathbf{p} \right]_{\s (\ell)}.
\end{equation}
Finally, the probability of another event occurring after time \(t\) can be obtained by another usage of the matrix \(\boldsymbol{\Gamma}\),
\begin{equation}\label{eq:trans-time-event}
    P(\lowprime{x}, t \vert x) = \mathcal{K}_{\lowprime{x}} \boldsymbol{1} \boldsymbol{\Gamma}_{\lowprime{x}} e^{t \mathbf{S}} \boldsymbol{\Gamma}_x \mathbf{p},
\end{equation}
with \(\boldsymbol{1}\) being a row vector of ones. See App.~\ref{app:proof} for a proof.

As proven in \cite{PhysRevE.107.L042105} and with an alternative approach in App.~\ref{app:inverse}, the survival matrix is invertible. Therefore, marginalizing the waiting-time leads to the probability of one event \(\lowprime{x}\) conditioned on the previous event \(x\):
\begin{equation}\label{eq:trans-event}
    P(\lowprime{x} \vert x) = - \mathcal{K}_{\lowprime{x}} \boldsymbol{1} \boldsymbol{\Gamma}_{\lowprime{x}} \mathbf{S}^{-1} \boldsymbol{\Gamma}_x \mathbf{p}.
\end{equation}
Although this equation is only conditioned on the last multifilar event \(x\), it does not mean the process is Markovian. The probability of a long sequence of events cannot be expressed by the multiplication of probabilities similar to Eq.~\eqref{eq:trans-event}.

Equations~\eqref{eq:trans-time-event} and \eqref{eq:trans-event} represent the statistics of pairs of multifilar events that will be used for thermodynamic inference in the following sections. They can be empirically estimated from experiments or numerical simulations. The transition-based coarse-graining framework is recovered from these equations by choosing an injective map \(\theta\). In this case, all matrices \(\Gamma\) are single-entry, \(\mathbf{p}_x\) has only a single entry 1, leading to Markovian sequences of observable transitions.

\subsection{Non-Markov}

The more general non-Markovian probability of a sequence of multifilar events can also be estimated using the introduced tools. The sequence of transitions and intertransition times, which is Markovian, is obtained by the path probability
\begin{equation}\label{eq:pathprob_transitions}
    P\left( \vec{\ell}, \vec{t}\, \right) = P(\ell_0) \prod_{i=1}^n P(\ell_i, t_{i} \vert \ell_{i-1}), 
\end{equation}
for \(\vec{\ell} = \{\ell_0,\ldots, \ell_n \}\) and \(\vec{t} = \{t_1,\ldots, t_n\}\). An expression analogous to Eq.~\eqref{eq:pathprob_transitions} for \(\vec{x}\) would represent the semi-Markov approximation. The path probability of multifilar events can be obtained from all possible sequences of observable transitions that map onto the sequence of events, i.e.,
\begin{equation}\label{eq:pathprob_events}
    P\left( \vec{x}, \vec{t}\, \right) =  \sum_{\vec{\ell} \in \mathcal{L}_\text{obs}^{n+1}} P\left( \vec{\ell}, \vec{t}\, \right) \prod_{i=0}^n \delta_{\theta \ell_i,x_i} ,
\end{equation}
where \(\delta\) is the Kronecker delta function.

Using the event-propagating matrix \(\boldsymbol{\Gamma}_x\) defined in Eq.~\eqref{eq:Gammamat}, this probability can be obtained by
\begin{equation}\label{eq:pathprob_nonmarkov}
    P\left( \vec{x},\vec{t}\, \vert x_0 \right) = \left( \prod_{i=1}^n \mathcal{K}_{x_i} \right) \mathbf{1} \left( \prod_{i=n}^1 \boldsymbol{\Gamma}_{x_i} e^{t_i \mathbf{S}} \right) \boldsymbol{\Gamma}_{x_0} \mathbf{p},
\end{equation}
where the product \(\prod_{i=n}^1\) is ordered with larger \(i\) on the left, i.e., \(\boldsymbol{\Gamma}_{x_n} \exp(t_n \mathbf{S}) \boldsymbol{\Gamma}_{x_{n-1}} \exp(t_{n-1} \mathbf{S}) \cdots \). This equation provides the joint probability density of a non-Markovian sequence of events \(\vec{x}\) and their waiting-times \(\vec{t}\). As in the previous case, this probability can be marginalized by integration of all waiting-times and the matrix exponentials will be replaced by \(- \mathbf{S}^{-1}\).

Equation~\eqref{eq:pathprob_nonmarkov} captures the path probability of each individual sequence of transitions that is compatible with the considered sequence of events. Intuitively, \(\exp(t \mathbf{S})\) captures the propagation through hidden pathways interspersed by multifilar events captured by \(\mathbf{\Gamma}\).

%
%

\section{Detecting and quantifying nonequilibrium}\label{sec:3}

The EPR of a continuous-time Markov chain is
\begin{equation}\label{eq:epr}
    \sigma = \sum_{ij} R_{ji} p_i \ln \frac{R_{ji} p_i}{R_{ij}p_j} = \mathcal{K} \sum_{\ell \in \mathcal{L}} P(\ell) \ln \frac{P(\ell)}{P(\bar{\ell})}
\end{equation}
in units of the Boltzmann constant~\cite{RevModPhys.48.571}. The quantities involved are transition rates and probability mass functions, and measuring EPR requires a method for their inference. Establishing a model involves the definition of the possible states and the transition rates between them, thus solving the master equation yields the probabilities and EPR is obtained. If no model is available, but all states can be observed, transition rates can be empirically obtained by the frequencies of each transition and probabilities are obtained from residence times. If there is a model but missing information, such as some inaccessible transition rates due to hidden states, many of the methods mentioned in Sec.~\ref{sec:introduction} can be used to estimate the EPR. Alternatively, it is also possible to devise model-free estimators that make no assumptions regarding the topology, existence of hidden states and their number, values of some transition rates, or more. These model-free estimators are invaluable for applying nonequilibrium thermodynamic methods to assess energetic interchanges, physical limitations, and the arrow of time in natural systems. When a model-free estimator is employed, its result is not conditioned on the validity of model assumptions. In the context of observing multifilar events, we discuss two model-free estimators and one that uses additional information to assess the distance to equilibrium; beyond accounting for multifilar events that cannot be resolved into individual transitions, all estimators can be applied in scenarios of partial information.


\subsection{Zero-knowledge}

The first model-free estimator considers the (unconditioned) probability of each event,
\begin{equation}\label{eq:Psingle}
    P(x) = \sum_{\ell \in \theta^{-1} x} P(\ell) = \sum_{\ell \in \theta^{-1} x} \frac{r_\ell p_{\mathrm{s}(\ell)}}{\mathcal{K}_\text{obs}},
\end{equation}
which can also be empirically obtained from frequencies.
With no requirements beyond the already announced assumptions, the log-sum inequality applied to Eq.~\eqref{eq:epr} establishes the lower bound
\begin{equation}\label{eq:sigma_zk}
    \sigma_\text{zk} \coloneqq \mathcal{K}_\text{obs} \sum_{x\in \mathcal{X}} P(x) \ln \frac{P(x)}{P(\bar{x})} \leq \sigma.
\end{equation}
If the multifilar events are associated with currents, it can be convenient to define \(j_x \coloneqq \mathcal{K}_\text{obs} [P(x) - P(\bar{x})]\) and 
\begin{equation}
    \sigma_\text{zk} = \frac{1}{2} \sum_{x\in \mathcal{X}} j_x \ln \frac{P(x)}{P(\bar{x})},
\end{equation}
where \(\ln P(x)/ P(\bar{x})\) acts as the effective affinity of each multifilar event.

Recall that the value of \(\mathcal{K}_\text{obs}\) is context-dependent, it is the traffic rate of all observable multifilar events. Therefore, it can always be measured by counting their number and dividing by the time span. The occurrence of hidden transitions does not contribute to its value.

With the estimator in Eq.~\eqref{eq:sigma_zk}, it is possible to bound the EPR from below using the statistics of each event in the absence of any knowledge about the exchanged quantities, the system topology, or physical characteristics such as the chemical potentials or temperatures of reservoirs. This quantity was sometimes referred to as a ``local'' entropy production and employed as an approximate measure. As will be illustrated, the bound can be tight even in the presence of \(\theta^{-1}x\) with infinite cardinality, as in chemical reaction networks.

For each pair \(\{ x, \bar{x} \}\), the sum of terms in Eq.~\eqref{eq:sigma_zk} is non-negative, and thus \(\sigma_\text{zk}\) still constitutes a lower bound if some events are hidden.


\subsection{Waiting-times}

When all transitions in a process are visible and distinguishable, waiting-time distributions generally do not play a role in entropy production~\cite{PhysRevX.12.041026, PhysRevX.12.031025}, since all are Poisson distributions of respective exit rates. The same is not valid for events due to their multifilar nature; thus, waiting-time distributions can entail additional thermodynamic information. Therefore, a very natural extension lies in considering the statistics of pairs of transitions, in which a semi-Markov approximation would be performed on the non-Markovian trajectories of events. Since it constitutes an approximation, the expressions may overestimate the real EPR, which is fundamentally a problem when the goal is to estimate thermodynamic costs and limits of a given process. Similarly, it has recently been suggested that waiting-times have to be disregarded in the semi-Markov approach \cite{blom2023milestoning}. We now show that waiting-time distributions between multifilar events establish a lower bound for EPR and serve as a nonequilibrium indicator.

Consider a subset \(\mathcal{L}_\text{obs}\) of all possible transitions, which is the set of visible transitions. When the occurrence of all elements of \(\mathcal{L}_\text{obs}\) is distinguishable (unifilar), the EPR is bound from below in the transition-based framework by
\begin{equation}\label{eq:PRX_sigma}
    \sigma \geq \mathcal{K}_\text{obs}\sum_{\ell', \ell \in \mathcal{L}_\text{obs}} \int_0^\infty \mathrm{d}t P(\ell, \ell', t) \ln \frac{P(\ell', t\vert \ell)}{P(\overline{\ell}, t \vert \overline{\ell'})},
\end{equation}
where \(P(\ell, \ell' ,t)\) is the joint probability density function for the occurrence of two consecutive transitions \(\ell\) and \(\ell'\) separated by a time interval \(t\)~\cite{PhysRevX.12.041026}. If the log-sum inequality is applied to make sums of multifilar events pop out, the result is
\begin{align}\label{eq:lowerbound}
    \sigma &
    \geq \mathcal{K}_\text{obs} \sum_{x, \lowprime{x} \in \mathcal{X}} \int_0^\infty \mathrm{d}t P(x, \lowprime{x}, t) \ln \frac{P(\lowprime{x},t \vert x)}{P(\overline{x} ,t\vert \overline{\lowprime{x}})} \nonumber\\
    &\phantom{\geq\ } + \mathcal{K}_\text{obs} \sum_{x \in \mathcal{X}} P(x) \ln \frac{P(x)}{P(\overline{x})} - \mathcal{K}_\text{obs} \sum_{\ell \in \mathcal{L}_\text{obs}}  P(\ell) \ln\frac{P(\ell)}{P(\overline{\ell})}.
\end{align}
Notice that the first term on the right-hand side only depends on the statistics of multifilar events, which we consider as the only available quantities, the second term is the zero-knowledge estimator., and the negative of the last term is smaller than or equal to the EPR. Therefore, we define a second-order estimator that leverages pairwise probabilities and waiting-time statistics to establish the lower bound:
\begin{equation}\label{eq:wt_bound}
    \sigma_\text{so} \coloneqq \frac{1}{2}\mathcal{K}_\text{obs} \sum_{x, \lowprime{x} \in \mathcal{X}} \int_0^\infty \mathrm{d}t P(x, \lowprime{x}, t) \ln \frac{P(x, \lowprime{x},t)}{P(  \overline{\lowprime{x}}, \overline{x} ,t)} \leq \sigma
\end{equation}
which was first proven by Ertel and Seifert in the coeval contribution~\cite{ertel2023estimator}. The inequality in Eq.~\eqref{eq:wt_bound} bounds the EPR from below using the available statistics of observable events and, importantly, accounts for their multifilarity.

Comparing the second-order with the zero-knowledge estimator, we see that \(\sigma_\text{so}\) has a contribution from waiting-times that is nonnegative due to its Kullback-Leibler divergence structure, thus any asymmetries in waiting-times can only contribute positively to the observable EPR. Furthermore, it is possible to show that
\begin{equation}
    \sigma_\text{so} \geq \sigma_\text{zk},
\end{equation}
even when waiting-times are fully symmetric, evidencing that the inclusion of conditional probabilities improves the estimation.

Beyond the estimation of EPR, the detection of nonequilibrium behavior is also valuable in many instances, it reveals a constant consumption of free energy and the presence of forces. Since the discussed estimators are nonnegative and smaller than or equal to the EPR, they vanish at thermal equilibrium, thus constituting nonequilibrium detectors. Additionally, it is convenient to decouple the contributions of the statistics of multifilar events and their waiting-times, since in some instances one of them might resemble equilibrium or be much more challenging to estimate from data, which is usually the case for the latter. Finally, we put forward two nonequilibrium detectors, stemming from the occurrence of events and their waiting-times:
\begin{equation}\label{eq:Dx}
    D_x \coloneqq \sum_{x,\lowprime{x} \in \mathcal{X}} P(x, \lowprime{x}) \ln \frac{P(x, \lowprime{x})}{P( \overline{\lowprime{x}}, \overline{x} )}
\end{equation}
and
\begin{equation}\label{eq:Dt}
    D_t \coloneqq \sum_{x,\lowprime{x} \in \mathcal{X}}  P(x, \lowprime{x}) D_{\text{KL}} [ P(t\vert x, \lowprime{x}) \vert\vert P(t\vert \overline{\lowprime{x}}, \overline{x}) ],
\end{equation}
where we denote as \(D_\text{KL}[P(t)\vert\vert Q(t)]\) the Kullback-Leibler divergence for continuous variables (see~\footnote{Empirical estimation of the divergence of continuous variables presents convergence problems, the reader may refer to the code in~\cite{Harunari_KLD_estimation} for an unbiased estimator.} for practical considerations) between \(P(t)\) and \(Q(t)\), which is the relative entropy of both distributions, and vanishes if and only if the distributions are the same. Equations~\eqref{eq:Dx} and \eqref{eq:Dt} are therefore relative entropies between a pair of events and its time-reversal analogue, with the former accounting for pairs of multifilar events and the latter for thei waiting-time between them. Both detectors are nonnegative and vanish at equilibrium, satisfying minimal conditions to be used as indicators of nonequilibrium behavior.

Importantly, the values of Eq.~\eqref{eq:Dx} and \eqref{eq:Dt} are nonnegative for any choice of \(\mathcal{X}\), and therefore are nonequilibrium detectors even if some events are hidden or there exist transitions not associated with events. Once again, we recall that if \(x \in \mathcal{X}\) then \(\bar{x} \in \mathcal{X}\). In summary, a positive value of \(D_x\) or \(D_t\) for any set of visible multifilar events \(\mathcal{X}\) is enough evidence for an underlying nonequilibrium process.

If the only observables are one event and its reverse, \(\mathcal{X} = \{x, \bar{x}\}\), irreversibility can be detected by comparing \((x,x)\) and \((\bar{x}, \bar{x})\), while the contribution from alternated terms \((x, \bar{x})\) and \( (\bar{x} ,x)\) vanish. Therefore, even if \(P(x) = P(\bar{x})\), which represents apparent equilibrium due to a stalled observable current, irreversibility can still be detected. Another simple example worth mentioning is the case of a system that only has two observable configurations \(\{0,1\}\) with many pathways connecting them. The sequence of events will be trivial: \(\{0 \to 1, 1 \to 0, 0 \to 1, \ldots \}\), indistinguishable from equilibrium behavior; however, the waiting-times can still reveal nonequilibrium by a nonvanishing \(D_t\).


\subsection{Topology-informed}

Additional information opens the possibility of analyzing the event statistics through a more informed lens and obtaining better bounds or estimators. We now consider the case of knowledge about the model in question and its state space topology, without the need for numerical values of transition rates or population probabilities. In general, this would not suffice to measure the EPR since it usually requires dynamical information in the form of transition rates.

A few classes of systems present a particular symmetry in the state space: more than one cycle sharing the same affinity, potentially an infinite number of them. In systems satisfying local detailed balance (LDB), which is a condition for thermodynamic consistency \cite{PhysRevE.85.041125, falascoLocalDetailedBalance2021a}, the affinity reduces to a sum over the entropy fluxes generated by each transition along the cycle that involves physical properties such as temperatures, energy exchanges, and chemical potentials, but not local state probabilities \footnote{Local detailed balance for a given transition reads \( r_\ell / r_{\overline{\ell}} = \exp(-\phi_\ell)\), with \(\phi_\ell\) the entropy flux generated by \(\ell\), and the affinity of a cycle becomes \(\sum_{\ell \in \vec{\mathcal{C}}} \phi_\ell\).}, and can result in the same value for families of cycles due to conservation laws \cite{polettiniConservationLawsSymmetries2016, Rao_2018}. Also, this is ubiquitous in systems driven by reservoirs, when the affinities are independent of local configurations and only expressed in terms of the intensive properties of reservoirs \cite{Rao_2018, falascoMacroscopicStochasticThermodynamics2023}. Notice that these are common mechanisms giving rise to said symmetry, but not necessary conditions. Examples include chemical reaction networks \cite{avanzini_methods_2023, 10__Schlogl1972}, stochastic models for electronic circuits \cite{freitasStochasticThermodynamicsNonlinear2021}, isothermal driven collective engines \cite{vroylandtCollectiveEffectsEnhancing2017}, and more. More specifically, in Section~\ref{sec:illustrations}, we explore in detail a double quantum-dot model \cite{PhysRevE.91.012145, Cuetara_2015, PhysRevE.103.032118} and a simplified Brusselator \cite{fritz_stochastic_2020, nicolis_self-organization_1977, qian2002concentration}, both of which fall into the said class. 

For the next estimator, we assume the following: There exists a finite number of families of cycles of the same affinity and, for each family, there is a unique minimal sequence of observable events \(\vec{x}_\alpha\) univocally defining the completion of a cycle \(\vec{\mathcal{C}}_\alpha\) from this family.

Network theory \cite{RevModPhys.48.571} says that the affinity of a cycle is given by the product of the rates of all edges involved, according to their convention orientation, 
\begin{equation}\label{eq:affinity_def}
    A_\alpha = \ln \frac{\prod_{\ell \in \vec{\mathcal{C}}_\alpha} r_\ell}{\prod_{\ell \in \cev{\mathcal{C}}_\alpha} r_\ell} .
\end{equation}
Next, we turn to the probability of observing a sequence \(\vec{x}_\alpha\), which is given by the sum of the probabilities of all transition sequences associated with \(\vec{x}_\alpha\):
\begin{equation}\label{eq:prob_sequence}
    P(\vec{x}_\alpha) = \sum_{\vec{\ell} : \theta \vec{\ell} = \vec{x}_\alpha}  P(\vec{\ell}\, ) = \sum_{\vec{\ell} : \theta \vec{\ell} = \vec{x}_\alpha} \frac{r_{\ell_0} p_{\s (\ell_0)}}{\mathcal{K}_\text{obs}} \prod_{i=1}^{\abs{\vec{\ell}\,} -1} \frac{r_{\ell_i}}{\mathrm{exit} ( \s (\ell_i))},
\end{equation}
where \(\mathrm{exit}( \bullet ) \coloneqq - [\mathbf{R}]_{\bullet, \bullet}  \) is the exit (or escape) rate of a state. Equation~\eqref{eq:prob_sequence} can be explicitly obtained using Eq.~\eqref{eq:pathprob_nonmarkov}. Since we assumed a one-to-one connection between a family of same-affinity cycles \(\vec{\mathcal{C}}_\alpha\) and a sequence of transitions \(\vec{x}_\alpha\), Eq.~\eqref{eq:affinity_def} can be plugged in the right-hand side of Eq.~\eqref{eq:prob_sequence}, making the factor \(\exp (A_\alpha)\) appear, and the log-ratio with the time-reversed sequence simplifies to
\begin{equation}\label{eq:affinity}
    \ln \frac{P(\vec{x}_\alpha)}{P(\cev{x}_\alpha)} = A_\alpha.
\end{equation}
\(\cev{x}_\alpha\) is defined as the time-reversal of the original sequence and can be obtained by reversing the order of the sequence and the direction of each individual event: \(x \leftrightarrow \overline{x}\). This time-reversal is nothing but the result of coarse-graining the time-reversed state space trajectory. For clarity, if \(\vec{x}_\alpha = \{ x_0, \ldots, x_n \}\), then \(\cev{x}_\alpha = \{ \bar{x}_n, \ldots, \bar{x}_0 \}\).

The entropy production can be expressed as a bilinear product of affinities and currents. If all affinities that contribute to the EPR are estimated using Eq.~\eqref{eq:affinity}, it is a matter of finding the cycle currents \(j_\alpha\). Depending on the model, they can be obtained from reservoir fluxes or by the flux over chords (edges removed from the network to define a spanning tree that forms the cycle in question when reintroduced). Given that the topology is known, defining this flux is most likely immediate. Bringing all together, we have the topology-informed estimator
\begin{equation}\label{eq:sigma_ti}
    \sigma_\text{ti} \coloneqq \sum_\alpha j_\alpha \ln \frac{P(\vec{x}_\alpha)}{P(\cev{x}_\alpha)} = \sigma,
\end{equation}
which is exactly the EPR. It is important to note that there exists some freedom in the definition of cycles, hence if the family of cycles that meet the requirements is not known, the following recipe can be followed for systems with finite state space: (i) choose a spanning tree (cf. \cite{RevModPhys.48.571}), (ii) identify cycles that share the same affinity and can be univocally defined by a sequence \(\vec{x}_\alpha\), (iii) measure the current at the chords of these cycles and the affinity using Eq.~\eqref{eq:affinity}. The method does not necessarily satisfy the assumption for every choice of spanning tree, so it might need to be repeated. For infinite state spaces, such as the concentration space of chemical reaction networks, the definition of a same-affinity cycle family usually comes from the list of reactions and chemostatted species.

We recall that the sequence of multifilar events is non-Markovian, the probabilities \(P(\vec{x}_\alpha)\) can be empirically obtained or calculated by the path probability of events, Eq.~\eqref{eq:pathprob_nonmarkov}. However, if a semi-Markov approximation is performed, the estimator can be employed to obtain an approximated value of the EPR that might be more feasible, since assessing the statistics of long sequences often becomes challenging.

For the topology-informed estimator \(\sigma_\text{ti}\), a partial sum of Eq.~\eqref{eq:sigma_ti} will not lead to an EPR lower bound since some terms might be negative, which is usually the case in heat engines, for example. Therefore, all families of cycles must be considered. Notice that not necessarily all system transitions are involved, thus the method can hold in partial information scenarios when the hidden transitions are not contained in \(\{ \vec{x}_\alpha \} \).
In other words, hidden transitions do not immediately rule out the assumption herein.


\subsection{Connection to previous results}

Before the establishment of the present results, the most appropriate alternative to estimate the EPR from partial observation of transitions is that of Refs.~\cite{PhysRevX.12.041026, PhysRevX.12.031025}, which form a strict lower bound for the unifilar case. If unifilarity cannot be assumed, the estimators therein would represent an approximation and not a strict lower bound. Although obtaining an approximation is a valid method, especially if the goal is to capture qualitative features, it is key to have a robust inequality in many tasks, such as in assessing energetics. Suppose an experimenter is observing the exchanges between a cell and its environment, it is crucial to obtain the minimal amount of \(k_\text{B}\) dissipated per time to keep its function rather than just an approximation; an overestimation might lead to incorrect assessment of properties such as energetic efficiency and the cost to sustain a process.

More precisely, the estimator for unifilar transitions is
\begin{equation}\label{eq:unifilar}
    \sigma_\text{unifilar} = \mathcal{K}_\text{obs} \sum_{\ell, \ell' \in \mathcal{L}} \int_0^\infty \mathrm{d}t P(\ell, \ell', t) \ln \frac{P(\ell', t \vert \ell)}{P(\bar{\ell} ,t \vert \bar{\ell'})},
\end{equation}
which is not equivalent to Eq.~\eqref{eq:wt_bound}. If multifilar events \(x\in\mathcal{X}\) are monitored and interpreted as unifilar transitions \(\ell\in\mathcal{L}\), regardless if as an approximation or a mistaken interpretation, and the values are fed into the estimator Eq.~\eqref{eq:unifilar}, this will no longer satisfy an inequality with the EPR and we will denote it \(\sigma_\text{approx}\) in this section.

\begin{figure}[t!]
    \centering
    \includegraphics[width=0.8\columnwidth]{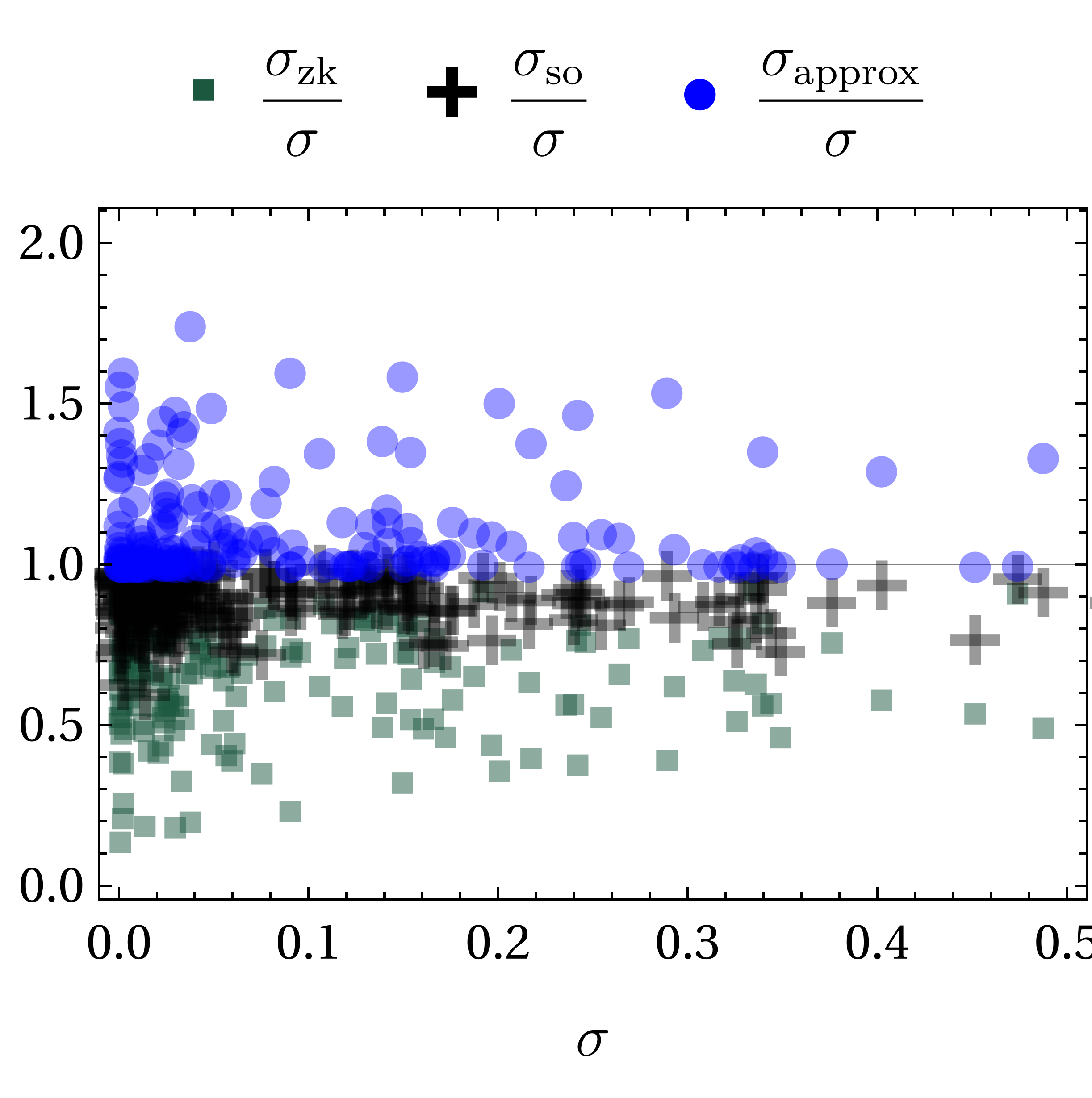}
    \caption{Distinct entropy production estimators in a unicyclic 4-state network, with observable multifilar events such that \(\theta^{-1} x_1 = \{ 3\to 4, 4\to 1 \}\) and \(\theta^{-1} x_2 = \{ 3\to 2\}\). All transition rates are randomized in \([0,1]\) for each dot.}
    \label{fig:comparison}
\end{figure}

Considering the minimal example of a unicyclic network with 4 states \(\{1,2,3,4\}\), we define one multifilar event \(x_1\) that occurs when transitions \(3\to 4\) or \(4\to 1\) occur, i.e. \(\theta^{-1} x_1 = \{ 3\to 4, 4\to 1 \}\), and also \(x_2\) such that \(\theta^{-1} x_2 = \{ 3\to 2\}\). Recall that if \(\theta \ell = x\), \(\theta \bar{\ell} = \bar{x}\). In Fig.~\ref{fig:comparison}, we plot the values of \(\sigma_\text{zk}\), \(\sigma_\text{so}\) and \(\sigma_\text{approx}\) normalized by the EPR for randomized rates in \([0,1]\). We observe that using \(\sigma_\text{approx}\) can indeed overestimate the EPR, while the other two estimators specialized for multifilar events are strict lower bounds.

\begin{figure*}[t!]
    \centering
    \includegraphics[width=\textwidth]{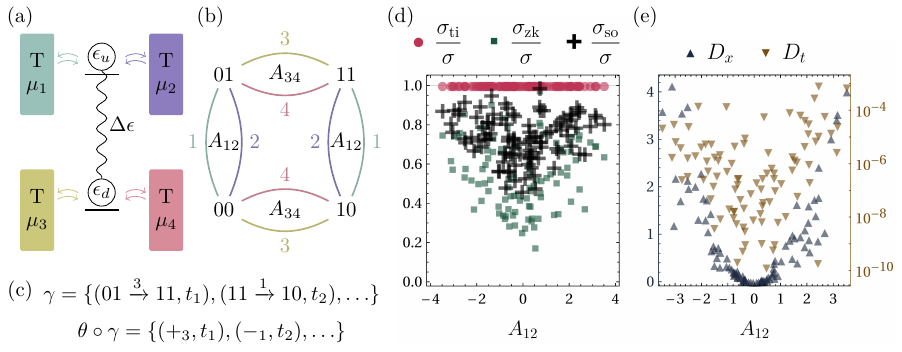}
    \caption{(a) Scheme of the double quantum-dot with four reservoirs and Coulomb repulsion; (b) state space with degenerate affinities \(A_{12}\) and \(A_{34}\) and the label/color of each edge represents the reservoir mediating the transitions and, therefore, its associated multifilar event observed from the reservoirs; (c) example trajectory $\gamma$ of electron exchanges and the respective observable trajectory $\theta \circ \gamma$ of multifilar events; (d) In terms of affinity \(A_{12}\) and normalized by the EPR, the exact estimator $\sigma_\text{ti}$, and the estimators $\sigma_\text{zk}$ and $\sigma_\text{so}$ evaluated from the events of reservoirs 1 to 3; (e) Nonequilibrium detectors $D_x$ (left axis) and $D_t$ (right axis) evaluated from the events of only reservoir 1. In the last two panels, $T=1$ and all other parameters are randomly chosen in $[1,5]$.}
    \label{fig:DQD}
\end{figure*}

%
%

\section{The case of reservoirs: a cautionary tale}

A quintessential problem in nonequilibrium thermodynamics consists of a system placed in contact with more than one reservoir with which it can exchange energy and/or matter. This is used to model electronic devices between leads, quantum transport problems, chemostatted chemical reactions, and the historical problem of a heat engine working between a hot and a cold reservoir. The events occurring at the reservoirs are usually multifilar and can be used to obtain entropy production through integrated currents if the physical quantities of all reservoirs are known. Let \(\mathbf{F}\) denote the fundamental forces, a set of nonconservative forces obtained by combining the physical properties of reservoirs and conservation laws~\cite{polettiniConservationLawsSymmetries2016, Rao_2018}, the EPR is
\begin{equation}
    \sigma = \mathbf{j} \cdot \mathbf{F}
\end{equation}
in the long-time limit and in units of the Boltzmann constant, where \(\mathbf{j}\) are physical currents associated with each force. Therefore, if fundamental forces are known, it is only required to measure values of currents, ruling out the need for inference schemes that employ, e.g. waiting-times analysis or semi-Markov approximations. It is worth pointing out that conservation laws might render some reservoirs futile, thus the number of fundamental forces can be smaller than the number of reservoirs, and the currents from futile reservoirs need not be monitored.

Fundamental forces are composed of quantities such as temperature, energy gaps, charges, and chemical potentials, but they may not be known or measurable with the desired accuracy. In these situations, specialized estimators provide tools to assess EPR or the presence of nonequilibrium behavior. The inverse problem is also relevant; once the EPR is estimated and the currents are known, something can be said about the fundamental forces. Now, we discuss how the results of previous sections can contribute to the case of observing events in reservoirs without knowing the thermodynamic forces.

In the case of reservoirs, the multifilar events are often increments \(\pm 1\) of a counter that monitors the number of exchanged quantities. If energy exchanges are also monitored, it might be possible to further resolve some events in terms of the energy they displace, enlarging the space of multifilar events and potentially resolving at the level of individual transitions and, naturally, improving the estimations. Without loss of generality, let's briefly
assume that all transitions between states are mediated by reservoirs. If some transitions are hidden, we imagine an extra reservoir mediating all of these transitions, so this scenario is always covered by the case of hidden reservoirs. The quantities \(\sigma_\text{zk}\), \(\sigma_\text{so}\), \(D_x\) and \(D_t\) are obtained as a sum of nonnegative terms, hence they can be used even when some reservoirs are hidden. Their values increase with the number of monitored reservoirs, thus the bounds due to \(\sigma_\text{zk}\) and \(\sigma_\text{so}\) get tighter, and it becomes increasingly easier to determine whether \(D_x\) or \(D_t\) are nonzero from finite data. Lastly, the estimator \(\sigma_\text{ti}\) requires visibility of all the reservoirs involved in the completion of cycles \(\vec{\mathcal{C}}_\alpha\), which does not necessarily include all reservoirs.

The relation between fundamental forces and affinities, if the local detailed balance condition is satisfied, is given by a ``\(M\)-matrix''~\cite{polettiniConservationLawsSymmetries2016} that captures the state space topology and the physical exchanges with each reservoir along all transitions. In some problems, it might be possible to relate the independent left-null vectors of this matrix and establish a direct connection between the affinity estimator in Eq.~\eqref{eq:affinity} and fundamental forces. However, we recall that, if the \(M\)-matrix is known, it is possible to obtain forces and currents analytically.

%
%

\section{Illustrations}\label{sec:illustrations}

\subsection{Double quantum-dot}\label{sec:dqd}

Motivated by \cite{PhysRevE.91.012145, Cuetara_2015, PhysRevE.103.032118}, we consider a model for two quantum-dots (two-level systems), each coupled to two reservoirs, whose interaction occurs by Coulomb repulsion. See Fig.~\ref{fig:DQD}(a) for the model scheme. We assume that all four reservoirs have the same temperature \(T = \beta^{-1}\) but different chemical potentials \(\mu_\nu\), the quantum-dots when occupied have energies \(\epsilon_\text{u}\) and \(\epsilon_\text{d}\), and the interaction between occupied dots carries energy \(\Delta \epsilon\). Transition rates satisfy LDB and thus the same affinity is present in distinct cycles as in Fig.~\ref{fig:DQD}(b).

If \(\mu_1 \neq \mu_2\) and \(\mu_3 \neq \mu_4\), the system is out of equilibrium and currents are established between each pair of reservoirs, passing through the quantum-dot between them. If \(\Delta \epsilon \neq 0\), both quantum-dots interact, and the currents are the result of every parameter of the system. Let \(\nu\) be the index representing each reservoir, we consider transition rates for \(\nu = \{1,2\}\)
\begin{align}
    R_{01,00}^{(\nu)} &= \Gamma_\nu f(\epsilon_\text{u} - \mu_\nu), \nonumber\\
    R_{00,01}^{(\nu)} &= \Gamma_\nu [1-f(\epsilon_\text{u} - \mu_\nu)], \nonumber\\
    R_{11,10}^{(\nu)} &= \Gamma_\nu f(\epsilon_\text{u} +\Delta\epsilon - \mu_\nu), \nonumber\\
    R_{10,11}^{(\nu)} &= \Gamma_\nu [1-f(\epsilon_\text{u} +\Delta\epsilon - \mu_\nu)],
\end{align}
where \(\Gamma_\nu\) is the coupling strength to the reservoir and \(f(\epsilon) = [1+\exp (\beta \epsilon)]^{-1}\) is the Fermi-Dirac distribution. For \(\nu = \{3,4\}\) they are
\begin{align}
    R_{10,00}^{(\nu)} &= \Gamma_\nu f(\epsilon_\text{d} - \mu_\nu), \nonumber\\ R_{00,10}^{(\nu)} &= \Gamma_\nu [1-f(\epsilon_\text{d} - \mu_\nu)], \nonumber\\
    R_{11,01}^{(\nu)} &= \Gamma_\nu f(\epsilon_\text{d} +\Delta\epsilon - \mu_\nu), \nonumber\\ R_{01,11}^{(\nu)} &= \Gamma_\nu [1-f(\epsilon_\text{d} +\Delta\epsilon - \mu_\nu)].
\end{align}

The state space presents many cycles and some share the same affinity due to LDB and isothermality. An analysis of the fundamental forces \cite{polettiniConservationLawsSymmetries2016, Rao_2018} reveals that only two affinities contribute to the EPR:
\begin{equation}\label{eq:sigma_dqd}
    \sigma = \beta (\mu_1 - \mu_2) J_1 + \beta (\mu_3 - \mu_4) J_3 = A_{12} J_1 + A_{34} J_2,
\end{equation}
where \(J_1 \coloneqq R_{01,00}^{(1)} p_{00} - R_{00,01}^{(1)} p_{01} + R_{11,10}^{(1)} p_{10} - R_{10,11}^{(1)} p_{11} \) is the flux from reservoir 1 t the system, and \(J_3 \coloneqq R_{10,00}^{(3)} p_{00} - R_{00,10}^{(3)} p_{10} + R_{11,01}^{(3)} p_{01} - R_{01,11}^{(3)} p_{11} \). These affinities are univocally related to the occurrence of some cycles, as highlighted in Fig.~\ref{fig:DQD}(b).

We consider electrons hopping into each reservoir as the observable multifilar events, which can be caused by distinct transitions and detected as increments in the voltage or a monitored current. For example, when a particle is provided by the first reservoir, we observe \(+_1\), and it can be similarly associated to either \(00 \xrightarrow{1} 01\) or \(10 \xrightarrow{1} 11\). See Fig.~\ref{fig:DQD}(c) for an example of a time series of this model and the associated multifilar events that can be observed from the reservoirs.

Figures~\ref{fig:DQD}(e) and (f) show the values of all quantities developed in Sec.~\ref{sec:3} evaluated for the double quantum-dot model, all dots represent exact values obtained using the analytical expressions in Sec.~\ref{sec:2}.

Notice that this model satisfies the condition of sequences of transitions univocally defining a family of cycles with the same affinity, they are \(\{+_1, -_2\}\) and \(\{+_3, -_4\}\), and the affinities are obtained by \(A_{12} = \ln [ P(+_1, -_2)/ P(+_2, -_1) ] = \beta (\mu_1 - \mu_2)\) and \(A_{34} = \ln [ P(+_3, -_4)/ P(+_4, -_3) ] = \beta (\mu_3 - \mu_4)\), hence the topology-informed method can be carried out to obtain these affinities and, consequently, the EPR even when the chemical potentials and temperature are not known. The currents are evaluated by \(J_1 = \mathcal{K} [P(+_1)- P(-_1)]\) and similarly for reservoir 3. The value of \(\sigma_\text{ti}\) is shown in Fig.~\ref{fig:DQD}(d), exactly obtaining the EPR from the statistics of events collected from the reservoirs.

The estimators \(\sigma_\text{zk}\) and \(\sigma_\text{so}\) are model-free, therefore is not necessary to identify cycles or make physical considerations about the processes involved, they can be obtained from the statistics of events regardless of what they are. In particular, Fig.~\ref{fig:DQD}(d) shows the case of observing when reservoirs 1, 2 and 3 provide or remove an  electron from the pair of quantum dots. We observe that indeed both are nonnegative and bound the EPR from below. The estimator \(\sigma_\text{zk}\) is interestingly tight, given that it only considers absolute probabilities of events that are not observed at the state space level and one of the reservoirs is hidden. Due to the inclusion of conditional probabilities and the contribution of waiting-times, the second-order estimator \(\sigma_\text{so}\) improves the estimation by making the bound tighter.

The nonequilibrium detectors \(D_x\) and \(D_t\) evaluated at reservoir 1 are shown in panel (e), with a positive value indicating the nonequilibrium character of the underlying process in state space. Notice that for \(A_{12} = 0\), the affinity between reservoirs 1 and 2 vanishes and their currents stall since no particles are exchanged between the two quantum-dots; hence the statistics of sequences of multifilar events collected at reservoir 1 is of apparent equilibrium and the detector \(D_x\) fails to detect nonequilibrium, even though \(A_{34}\) is not necessarily zero. However, due to the repulsive interaction and the non-zero affinity \(A_{34}\), the waiting-time distributions become asymmetric and \(D_t\) is able to reveal nonequilibrium using only the statistics of reservoir 1. The values shown in Fig.~\ref{fig:DQD} are exact and, even though \(D_t\) is usually small, it is possible to decide whether they are zero or not by looking at the fluctuations with respect to many realizations of the process.

\begin{figure}[t!]
    \centering
    \includegraphics[width=.75\columnwidth]{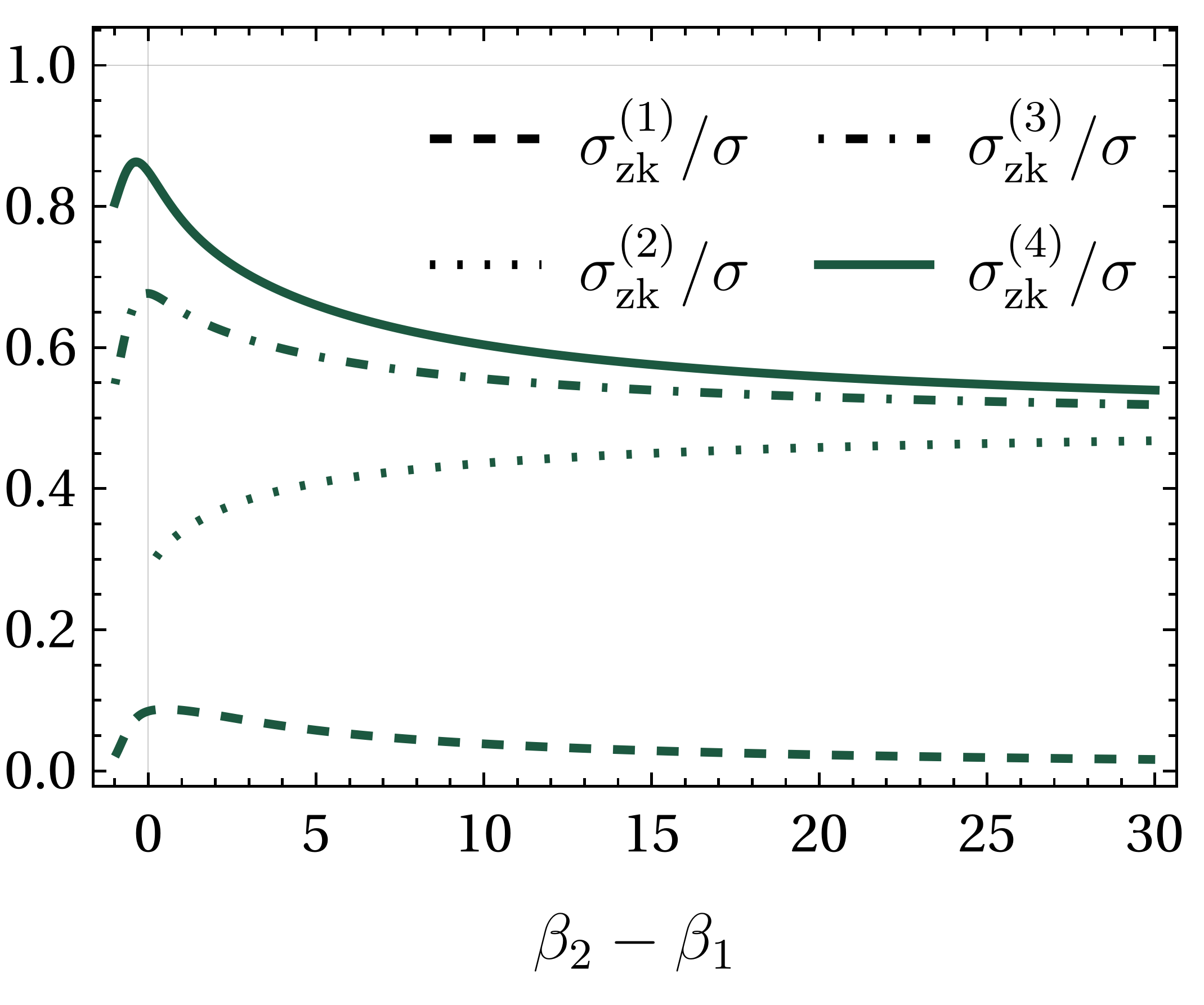}
    \caption{Zero-knowledge estimator \(\sigma_\text{zk}^{(n)}\) for the observation of the \(n\) first reservoirs. Inverse temperature \(\beta_2\) has values in \([0,31]\), and the other parameters are \(\{\Gamma_1, \Gamma_2, \Gamma_3, \Gamma_4\} = \{11, 5, 7, 13\}\), \( \{\mu_1, \mu_2, \mu_3, \mu_4\} = \{5, 1, 6, 1\}\), \(\beta_{1} = \beta_3 = \beta_4 = 1\) and \(\{ \epsilon_\text{u}, \epsilon_\text{d}, \Delta\epsilon\} = \{2,1,3\}\).}
    \label{fig:DQD_noniso}
\end{figure}

\begin{figure*}[t!]
    \centering
    \includegraphics[width=\textwidth]{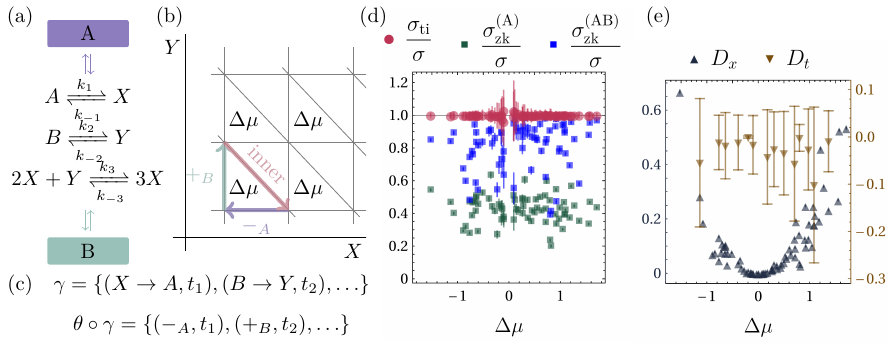}
    \caption{(a) Reactions of the considered Brusselator model and reservoirs chemostating species \(A\) and \(B\); (b) state space with degenerate affinities \(\Delta\mu\) and highlighted multifilar events that can be observed from the chemostats and the internal event; (c) example trajectory $\gamma$ of chemical reactions and the respective observable trajectory $\theta \circ \gamma$ of multifilar events; (d) In terms of affinity \(\Delta\mu\) and normalized by the EPR, the exact estimator $\sigma_\text{ti}$, the estimator $\sigma_\text{zk}^{(A)}$ evaluated from reservoir \(A\), and $\sigma_\text{zk}^{(A)}$ from reservoirs \(A\) and \(B\); (e) Nonequilibrium detectors $D_x$ and $D_t$ evaluated from the events of only reservoir \(A\). In the simulations, $V=5$, $[A]$ and $[B]$ are randomly drawn in $[5,10]$, and the rates $k_{\pm i}$ in $[1,3]$; each dot represents the average value of the estimator evaluated from 100 trajectories of $t_\text{max} = 10^4$ and error bars are shown.}
    \label{fig:Bruss}
\end{figure*}

In the case of reservoirs with different temperatures, the non-isothermal scenario, the affinities will not be the same throughout the distinct cycles and the method of \(\sigma_\text{ti}\) does not apply. For instance, the cycle between 00 and 01 will have affinity \(\epsilon_u (\beta_2 - \beta_1) +\beta_1 \mu_1 - \beta_2 \mu_2\), while between 10 and 11 it is \( (\epsilon_u + \Delta\epsilon) (\beta_2 - \beta_1) +\beta_1 \mu_1 - \beta_2 \mu_2\). It is evident that the same affinity case is recovered when \(\beta_1 = \beta_2\). Each affinity cannot be measured since the occurrence of individual cycles cannot be resolved from the multifilar events of particle exchanges, although it would be possible if energy transfers were also monitored. Also, there is an affinity \(\Delta\epsilon (\beta_1 - \beta_3)\) related to the cycle \(00 \xrightarrow{1} 01 \xrightarrow{3} 11 \xrightarrow{1} 10 \xrightarrow{3} 00\) that contributes to the EPR and cannot be obtained from multifilar events. Figure~\ref{fig:DQD_noniso} shows the EPR lower bound using the zero-knowledge estimator. Changing one of the temperatures has a nontrivial influence on the estimators, but the main features are preserved: It is still a lower bound and the addition of reservoirs makes it tighter.

In the presence of transitions not mediated by any of the reservoirs, such as in the possibility of electron hops between quantum-dots \(01 \leftrightarrow 10\), the estimators \(\sigma_\text{zk}\) and \(\sigma_\text{so}\) would still provide a lower bound, and the detectors \(D_x\) and \(D_t\) would still be nonnegative and vanish at equilibrium. The sequence of some particular events would still measure the affinities using Eq.~\eqref{eq:affinity}, however the currents will change and there will be an additional contribution to the EPR, thus considering a partial set of affinities in \(\sigma_\text{ti}\) does not provide the exact EPR and might even overestimate it, witnessing the need to be aware of all the affinities at play when using this estimator.

\subsection{Brusselator}

The second model considered is composed of three chemical reactions, two fluctuating chemical species \(X\) and \(Y\), and two species whose concentrations are fixed by chemostats [see Fig.~\ref{fig:Bruss}(a)]. It is a simplified version of the Brusselator model~
\cite{fritz_stochastic_2020, nicolis_self-organization_1977, qian2002concentration} that is largely studied due to its nontrivial behavior, such as the presence of a limit cycle and chemical oscillations emerging through a phase transition \cite{qian2002concentration, nguyen2018phase}. The state space is the infinite 2D lattice of concentrations of \(X\) and \(Y\) with reactions translating into horizontal (\(k_{\pm 1}\)), vertical (\(k_{\pm 2}\)), and diagonal (\(k_{\pm 3}\)) transitions. The affinity is given by \(\Delta\mu = \ln ( [B] k_2 k_{-1} k_3 / [A] k_{-2} k_1 k_{-3})\) and the entropy production is \(\sigma = J \Delta \mu\), with \(J\) being the current from reservoir \(B\) to the system (or minus the current from \(A\) to the system). The kinetic constants \(k_{\pm \rho}\) are translated into transition rates according to the rule of mass-action kinetics:
\begin{align}
    r_1 = k_1 [A] &\qquad r_{-1} = k_{-1} [X] \nonumber\\
    r_2 = k_2 [B] &\qquad r_{-2} = k_{-2} [Y] \nonumber\\
    r_3 = k_3 \frac{[X][X-1][Y]}{V^2}  &\qquad r_{-3} = k_{-3} \frac{[X][X-1][X-2]}{V^2}
\end{align}
where \(V\) is the volume of the vessel where reactions take place.

As seen in Fig.~\ref{fig:Bruss}(b), the state space has an infinite number of cycles, all with affinity \(\Delta\mu\) that can be univocally related to the sequence of reactions \(\{-_A, +_B, +_I\}\). The first two types of reactions give rise to multifilar events that can be monitored as exchanges of chemostatted species between system and reservoir. If we assume that the internal reactions \(\pm_I\) are also monitored, then the affinity can be empirically estimated by \( \ln P(-_A, +_B, +_I) / P(-_I, -_B, +_A)\); then the EPR is obtained by multiplying the estimated affinity by the current from one of the reservoirs to the system. This result is shown in Fig.~\ref{fig:Bruss}(d) through numerical simulations for randomized parameters and in agreement with the exact EPR. For larger values of \(\Delta\mu\), the error bars are smaller and the inference more precise. Close to equilibrium, the error bars get larger since it is harder to pinpoint the exact value when the EPR itself is very small; for aesthetic purposes, we have removed the points of smallest \(\Delta\mu\) since the error bars can overextend the size of the plot. This discussion considers that the internal transitions are visible but, in general, they do not give rise to observables, thus we now turn to the other estimators.

When the only observables are changes in the chemostats, the zero-knowledge estimator \(\sigma_\text{zk}\) and the second-order estimator \(\sigma_\text{so}\) can be used to bound the EPR from below. The former is shown in Fig.~\ref{fig:Bruss}(d) for the probabilities of events observed at reservoir \(A\) through \(\sigma_\text{zk}^{(A)}\), estimating about half the EPR for distinct sets of parameters. When reservoir \(B\) is also observed, the estimator \(\sigma_\text{zk}^{(AB)}\) yields an even tighter bound for the EPR. We remind that they require no knowledge about rates \(k_{\pm \rho}\) or the chemostatted concentrations \([A]\) and \([B]\), and they also do not involve assessment of the affinity \(\Delta\mu\), these bounds only concern the promptly accessible statistics of changes in the chemostats and directly measure the irreversibility of individual multifilar events.

Since \(\Delta\mu\) is the only affinity of the model, the system is in equilibrium when it vanishes and all estimators/detectors vanish for \(\Delta\mu = 0\). Figure~\ref{fig:Bruss}(d) shows the nonequilibrium detectors \(D_x\) and \(D_t\) evaluated from the statistics of chemostat \(A\) only. The values of \(D_x\) indicate the presence of nonequilibrium behavior when \(\Delta\mu \neq 0\) and vanish otherwise. Interestingly, its waiting-times counterpart \(D_t\) fails in detecting nonequilibrium behavior even in the presence of a nonzero affinity, being always compatible with zero with large error bars for the same simulations that are enough for the other estimators. Due to this behavior of the waiting-time distributions, \(\sigma_\text{so}\) does not represent an improvement over \(\sigma_\text{zk}\) and has error bars too large to contribute to the estimation.

\subsection{A paradigmatic 4-state system}

One common scenario that gives rise to multifilar events is the lumping of states, where all transitions between two lumped states become indistinguishable. For instance, this scenario can arise in the double-quantum dot of Section~\ref{sec:dqd} when the charge is measured, but the detector cannot determine which quantum dot is occupied if the charge is compatible with a single electron. In this case, the measurement can only inform if there are 0, 1 or 2 electrons inside the system. Other examples are a ligand that binds to a macromolecule while the specific receptor remains unresolved \cite{fernandesmartinsTopologicallyConstrainedFluctuations2023}, and a passive mode experiment of two DNA hairpins in series, which has the potential to be used as a Szilard engine \cite{schonenbergerFoldingEnergyKinetics}.

\begin{figure}[t!]
    \centering
    \includegraphics[width=0.8\columnwidth]{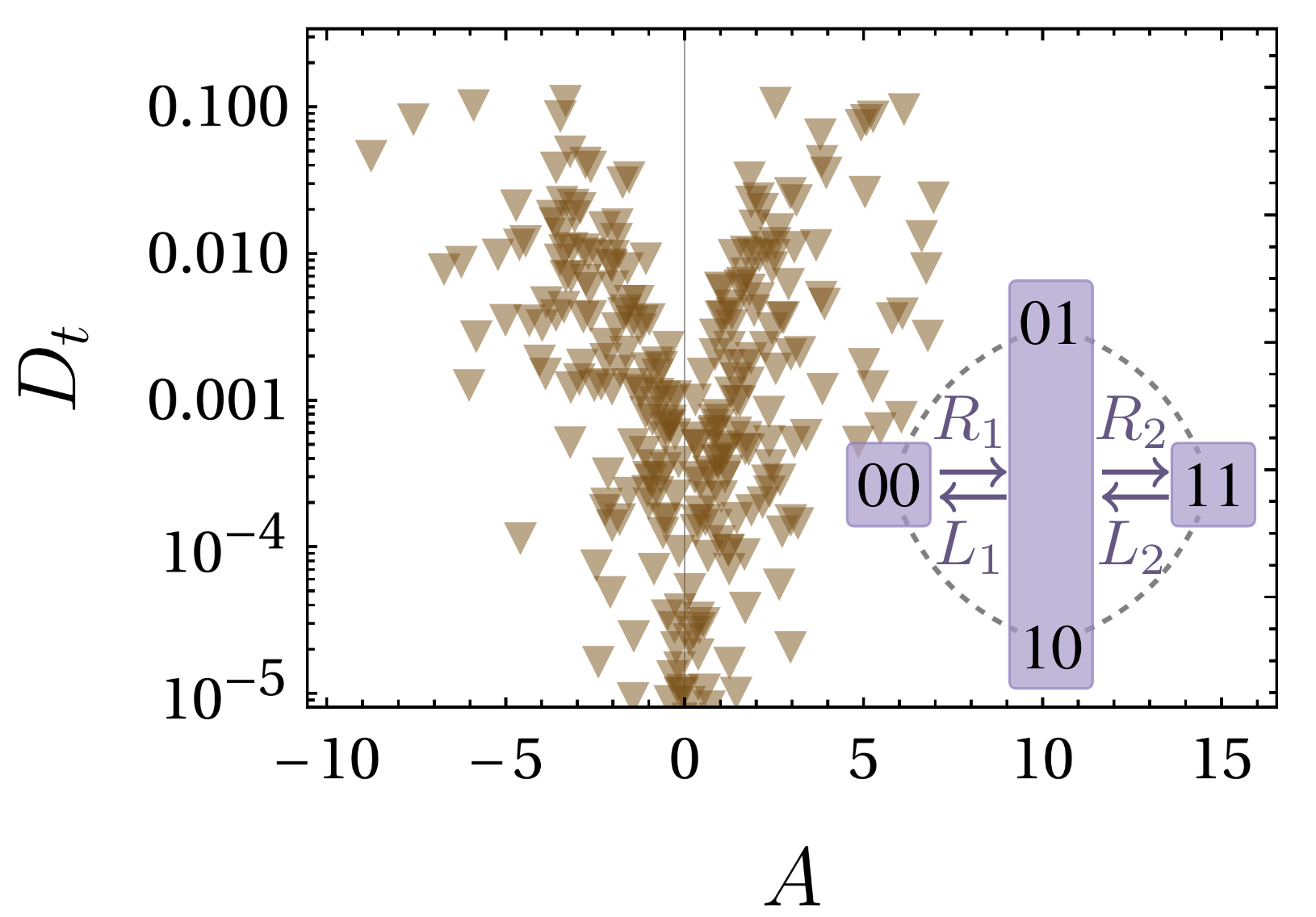}
    \caption{Detection of nonequilibrium from waiting-time distributions via $D_t$ for a 4-state model where two states are indistinguishable, hence the observable multifilar events are $\theta^{-1}R_1 = \{ 00 \to 01, 00 \to 10\}$, $\theta^{-1}R_2 = \{ 10 \to 11, 01 \to 11\}$, $L_1 = \bar{R_1}$ and $L_2 = \bar{R_2}$. All transition rates are randomized in \([0,3]\) for each marker and $A$ is the affinity in the clockwise direction.}
    \label{fig:4state}
\end{figure}

Inspired by these examples, we explore a model with states $\{ 00, 01, 10, 11\}$ and consider the scenario of $01$ and $10$ being indistinguishable, which is also a minimal model to study the effects of state lumping. In this case, the transitions $00 \to 01$ and $00 \to 10$ give rise to the same multifilar event $R_1$, and their time reversal yields $L_1$. Similarly, $01 \to 11$ and $10 \to 11$ are associated with $R_2$. See the inset of Fig.~\ref{fig:4state}. The single affinity $A$ of this system can be evaluated as the natural logarithm of the product of rates in the clockwise direction divided by the product of those in the counterclockwise direction.

Since probability is conserved, there cannot be a net flux of probability from state $00$ to $\{01, 10\}$, therefore $P(R_1) = P(L_1)$. This property can also be seen as the vanishing flux over a cocycle \cite{Cengio2022}. Similarly, $P(R_2) = P(L_2)$. Notice that this does not imply that the system is at thermal equilibrium since there can be a net flux over $00 \leftrightarrow 10$ and so on. This equivalence of multifilar event probabilities implies that $\sigma_\text{zk}$ is always zero and does not suffice to estimate/detect nonequilibrium.

Further, the joint probabilities $P(R_1, R_2)$ and $P(L_2, L_1)$ are equal again by conservation of probability. Therefore, $D_x$ also vanishes and the only contribution to $\sigma_\text{so}$ is due to the waiting-time distributions. The statistics of individual events or pairs in $\mathcal{X} = \{R_1, L_1, R_2, L_2\}$ are always compatible with equilibrium and, for the purpose of detecting nonequilibrium behavior, the detector $D_t$ emerges as the only resource from the tools developed herein. Figure~\ref{fig:4state} illustrates the performance of this detector through analytical calculations, which can reach $10^{-1}$ for higher affinity values.

%
%

\section{Discussion}

In summary, we extend the framework of transition-based coarse-graining to the case of multifilar events, which represents a large class of systems and is commonplace in experiments. We also develop methods to detect and quantify nonequilibrium behavior from the statistics of multifilar events: the zero-knowledge estimator that uses the probabilities of each event, the second-order estimator that also includes conditional probabilities and waiting-time distributions, the topology-informed that considers model and topological information to exactly obtain the affinities and consequently the full EPR, and two detectors of nonequilibrium behavior. Importantly, \(\sigma_\text{zk}\), \(\sigma_\text{so}\), \(D_x\) and \(D_t\) are model-free quantities, they can be applied to sequences of multifilar events of any nature and do not require further assumptions. Similarly, \(\sigma_\text{ti}\) uses topological and model information, bypassing the need to know or estimate transition rates, chemical potentials, or any other quantities that are often difficult to measure. Therefore, they are relevant for bridging experiments, where partial information is the rule, to the toolbox of nonequilibrium thermodynamics. Similarly, they can be used to validate proposed models.

We have seen that \(D_t\) can detect nonequilibrium behavior through asymmetries in waiting-time distributions even when the current of the monitored reservoir vanishes, at the same time this estimator is always compatible with zero for the simplified Brusselator model, a behavior that also renders \(\sigma_\text{so}\) unreliable when compared to \(\sigma_\text{zk}\). It would be interesting to understand the mechanisms behind its behavior and the classes of systems in which \(D_t\) becomes ineffective, which is possibly related to the number of chemical reactions, the topology of its state space, or properties such as deficiency and emergent cycles.

As illustrated by the double quantum-dot model, only two reservoir currents must be monitored if the affinities are known but, when they are not, \(\sigma_\text{zk}\) becomes tighter when more than two reservoirs are monitored. This is due to the lack of physical information in this bound. Another possible extension of the present results is to account for possible physical symmetries owing to known conservation laws. Also, to move from the detection of cycle affinities to thermodynamic forces, to understand the role of conserved quantities in identifying detectable families of cycles, and to explore the case of continuous multifilar events where position or energy exchanges are detected under limited resolution.

Further examples of setups where the present results can be relevant include pulling experiments involving more than two DNA hairpins in series, emission of photons without detecting which atoms are in the excited state, observed RNA elongation from unresolved transcription loci, molecular motors with many mechanisms driving the mechanical steps, more convoluted chemical reaction networks, and an unknown electronic device between two monitored leads. Also, in setups where there is no assumed Markov model or the model is under question, employing the developed model-free estimators can reveal otherwise hidden properties of the system from experimentally accessible data.

\section{Data Availability Statement}

The codes and data used to generate the analytical and numerical results of the present work are available and described in the repository~\cite{github_multifilar}.

\section{Acknowledgments}

We thank Massimiliano Esposito for useful discussions. This research was supported by the project INTER/FNRS/20/15074473 funded by F.R.S.-FNRS (Belgium) and FNR (Luxembourg).

\appendix

%
%

\section{Auxiliary process with multifilarity}\label{app:proof}

The joint probability density of the next multifilar event and its waiting-time, conditioned on the previous event, can be obtained by solving a first-transition time problem. The solution is a simple extension of the proof in Appendix A of Ref.~\cite{PhysRevX.12.041026}, but the matrix \(\mathbf{L}\) defined therein now has the contribution of more than one transition per element. The proof sketch is similar, and we present it here for consistency.

For a Markov process with generator \(\mathbf{R}\), we define multifilar events \(x\) that immediately occur when a transition \(\ell \in \theta^{-1} x\) is performed. The auxiliary process is established by the introduction of an absorbing state for each event, and all transitions that generate these events are redirected towards its respective absorbing state~\cite{sekimoto2021derivation}. The generator of the auxiliary process is the block matrix
\begin{equation}
    \mathbf{R}_\text{aux} = \left( \begin{array}{c@{}|c@{}}
    \mathbf{S} \hspace{2pt} & 0 \\
    \hline
    \mathbf{L} \hspace{2pt} & 0 \\
\end{array} \right),
\end{equation}
where \(\mathbf{S}\) is the survival matrix in Eq.~\eqref{eq:survivalmatrix} and 
\begin{equation}
    [\mathbf{L}]_{x,j} \coloneqq \sum_{\ell \in \theta^{-1} x} r_\ell \delta_{j, \s(\ell)}
\end{equation}
with each row \(x\) representing one of the events. Notice that their order is not relevant. The formal solution of the master equation, \(\mathbf{p}(t) = \exp(t \mathbf{R} ) \mathbf{p}(0)\), requires the matrix exponential of the generator, the propagator, that for the auxiliary dynamics reads
\begin{equation}\label{eq:exponential_auxiliary}
    \exp (t \mathbf{R}_\text{aux}) = \left( \begin{array}{c@{}|c@{}}
    \exp (t \mathbf{S}) \hspace{2pt} & 0 \\
    \hline
    \mathbf{L} \mathbf{S}^{-1} [\exp (t \mathbf{S} -1)] \hspace{2pt} & 0 \\
\end{array} \right).
\end{equation}

The probability that the process evolves during time \(t\) without reaching a particular absorbing state associated with \(x\) is known as its survival probability. Conditioning on a particular initial distribution \(\mathbf{p}_\text{aux}^0\), the survival probability is
\begin{equation}
    \mathbb{S} (t,x \vert \mathbf{p}_\text{aux}^0) = 1 - [\exp (t \mathbf{R}_\text{aux}) \mathbf{p}_\text{aux}^0]_x.
\end{equation}
Assuming that the initial distribution has entries zero in the rows associated with sinks, which is the relevant case here, it is a vector with first entries \(\mathbf{p}^0\) that is a valid distribution in the original state space. Hence, the only elements that will be relevant come from the bottom-left block of Eq.~\eqref{eq:exponential_auxiliary}. The first-passage time is obtained as minus the time derivative of the survival probability, and the joint probability we aim at is precisely the first-passage time distribution, therefore
\begin{equation}
    P(x, t\vert \mathbf{p}_\text{aux}^0) = \sum_{\ell \in \theta^{-1} x} r_\ell [\exp( t \mathbf{S}) \mathbf{p}^0]_{\s(\ell)}.
\end{equation}

Now, if the process is conditioned on the previous multifilar event, \(\mathbf{p}^0 \to \mathbf{p}_{\vert x}\). Finally, the joint probability density to perform an event \(\lowprime{x}\) after waiting-time \(t\) becomes equivalent to Eq.~\eqref{eq:trans-time-event}.

%
%

\section{Invertibility and convergence of the auxiliary process}
\label{app:inverse}

Here, we prove that the propagator of the survival process \(\exp ( t \mathbf{S})\) is finite at long times, and that \(\mathbf{S}^{-1}\) exists. First, we note that, since probabilities have to be preserved, \(\mathbf{1} \cdot \mathbf{p} = 1\), the columns of a continuous-time Markov chain generator \(\mathbf{R}\) sum to zero. It can be seen as a property of the master equation's formal solution; consider a small enough \(\delta t\), then
\begin{align}
    \mathbf{1}\cdot \left[\mathbf{p} (t + \delta t)\right] &= \mathbf{1}\cdot \left[e^{\delta t \mathbf{R}} \mathbf{p} (t) \right] \nonumber\\
    &= \mathbf{1}\cdot \left[ \left(\mathbf{I} + \delta t \mathbf{R} + \mathcal{O}[\delta t^2] \right) \mathbf{p} (t) \right] \nonumber\\
    &= 1+ \delta t \left( \mathbf{1} \cdot \mathbf{R}\right) \mathbf{p} (t) = 1,
\end{align}
where we dropped the higher-order contributions.
Second, if the process is supported by an irreducible network, the generator \(\mathbf{R}\) has a non-degenerate eigenvalue equal to zero, and its respective eigenvector has non-zero entries describing the stationary distribution. This is a consequence of applying the Perron-Frobenius theorem to a discretized version of the Markov chain.

The Gershgorin circle theorem \cite{gershgorin1931uber} states that all eigenvalues \(\lambda\) of a matrix \(\mathbf{R}\) lie within at least one disc \(D_i(\mathbf{R}) \subseteq \mathbb{C}\) in the complex plane, where each disk is centered at the diagonal element \(R_{ii}\) and has a radius equal to the sum of the non-diagonal elements \(\sum_{j \neq i} R_{ji}\). Since every diagonal element of a generator is equal to minus the sum of non-diagonal entries in its column, every \(D_i( \mathbf{R})\) is contained in the non-positive side of the real axis. Therefore, the real part of all eigenvalues is non-positive, in agreement with Perron-Frobenius.

\begin{figure}
    \centering
    \includegraphics[height = 130pt]{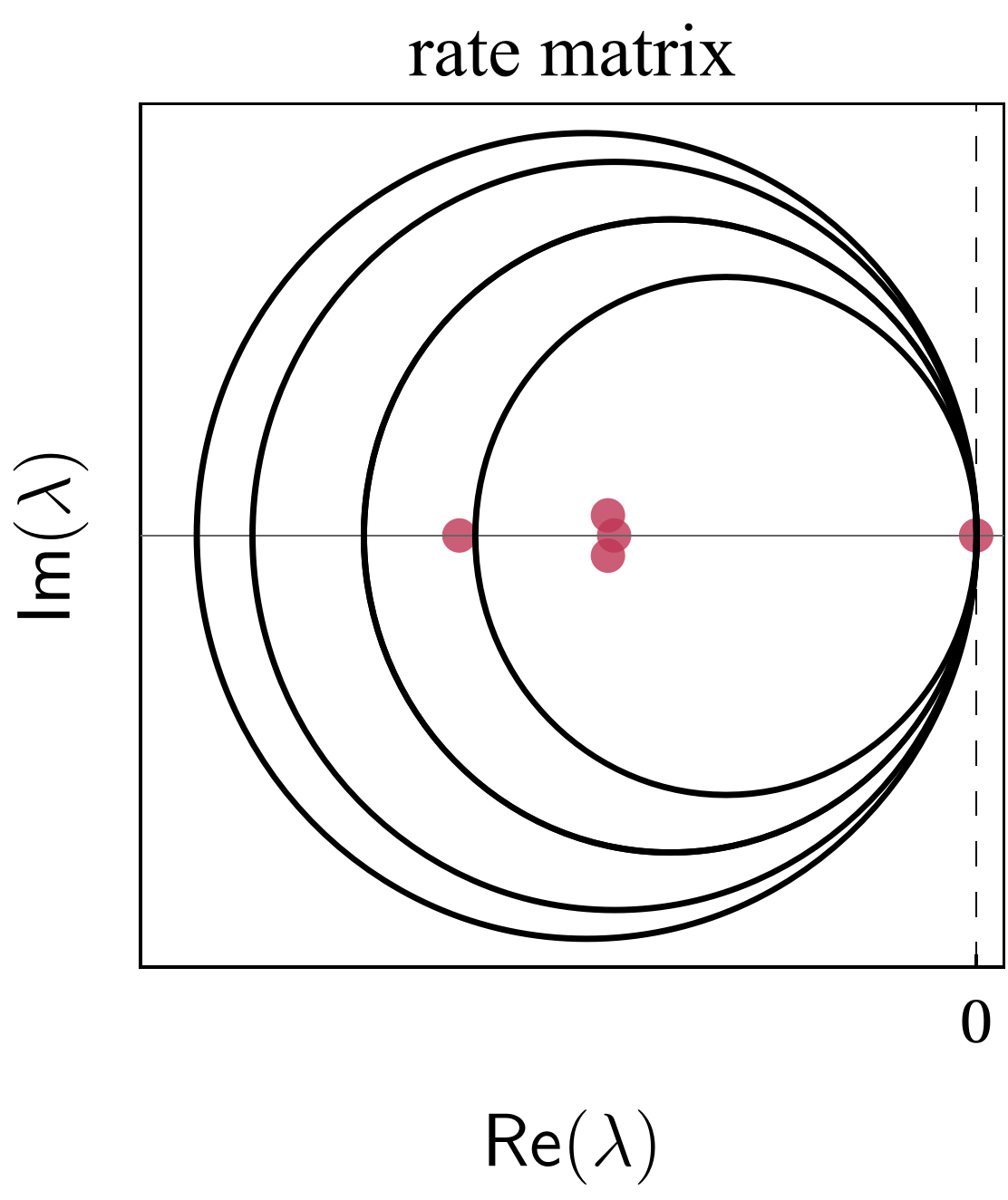}%
    \hspace{.01\columnwidth}%
    \includegraphics[height = 130pt, clip, trim = 20pt 0 0 0]{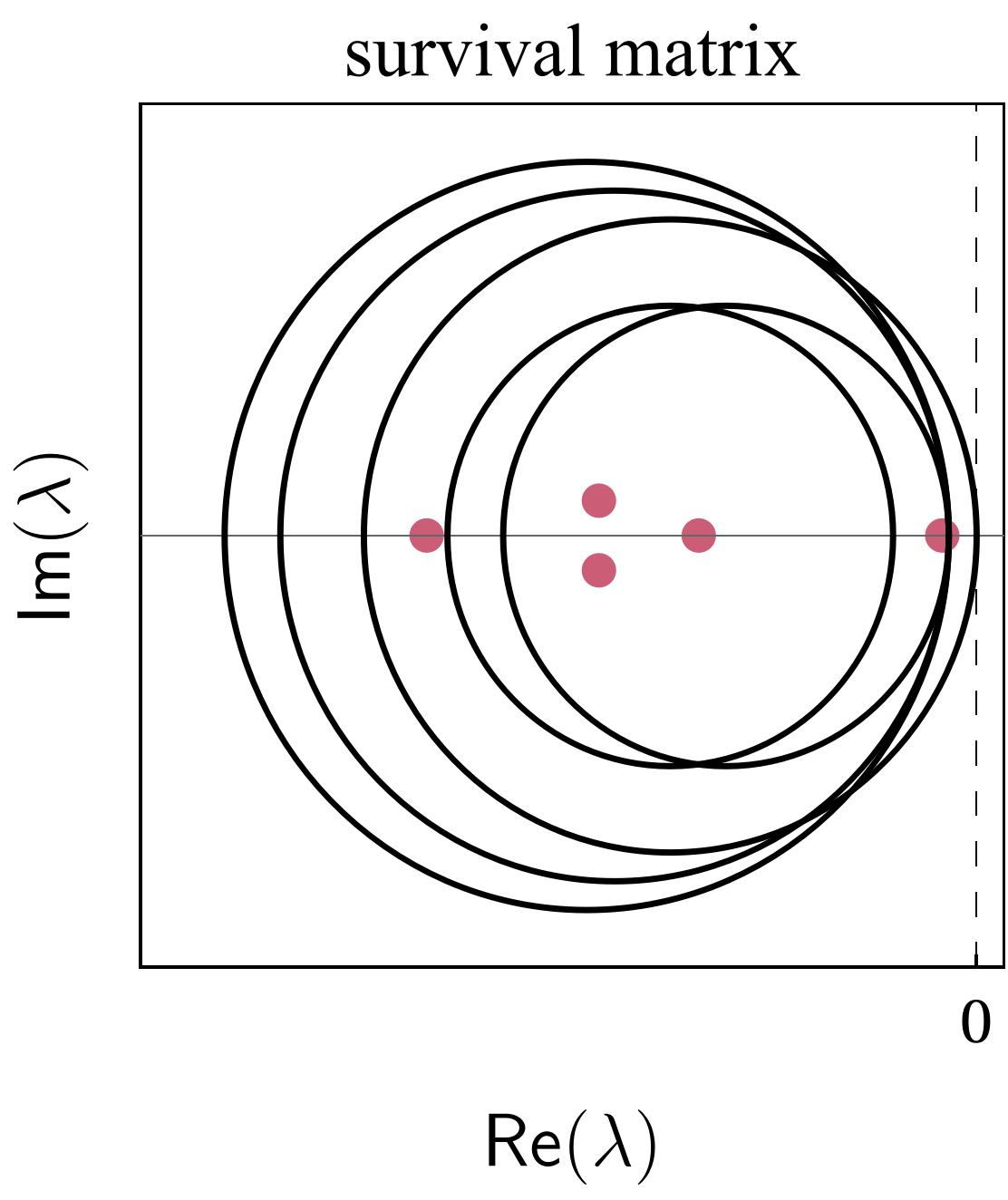}
    \caption{Gershgorin discs (black) and eigenvalues (red dots) for the rate matrix \(\mathbf{R}\) (left) and the survival matrix \(\mathbf{S}\) (right). The state space has 5 states and is fully connected, to construct \(\mathbf{S}\) we consider two visible transitions.}
    \label{fig:gershgorin}
\end{figure}

By construction, matrices \(\mathbf{R}\) and \(\mathbf{S}\) share the same diagonal elements (exit rates), but the columns of \(\mathbf{S}\) will sum to a value smaller than or equal to the sum of those of \(\mathbf{R}\). Therefore, one or more of the Gershgorin discs \(D_i(\mathbf{S})\) are shrunk versions of \(D_i(\mathbf{R})\), see Fig.~\ref{fig:gershgorin}. Hence the real part of all eigenvalues of \(\mathbf{S}\) are also non-negative.

Notice that the eigenvalues are functions of the matrix elements. The process of constructing \(\mathbf{S}\) from \(\mathbf{R}\) involves reducing some of its values and therefore changing the eigenvalues. Since the new eigenvalues have to fall in one of the Gershgorin discs, which are all in the non-positive side, the eigenvalue with the largest real part will acquire a negative real part, which is a sufficient condition for the invertibility. It is worth commenting that there are characteristic polynomial-preserving transformations, such as swapping rows or columns, which are not involved in the survival matrix. In addition, similarity transformations would leave the eigenvalues unchanged, so if there exists an invertible \(\mathbf{P}\) such that \(\mathbf{S} = \mathbf{P}^{-1} \mathbf{R} \mathbf{P}\), then \(\mathbf{S}\) will not be invertible. However, it would require a very particular symmetry to achieve such a similarity relation, which is unlikely to be possible under the condition of irreducibility.

Finally, Sylvester's formula states that for a diagonalizable matrix \(\mathbf{S}\)
\begin{equation}\label{eq:etS}
    \lim_{t\to\infty} e^{t \mathbf{S}} = \lim_{t\to\infty}  \sum_k A_k e^{t \lambda_k} = \lim_{t\to\infty} A_{\tilde{k}} e^{t \lambda_{\tilde{k}}},
\end{equation}
where the sum spans through all eigenvalues and \(\tilde{k}\) indexes the eigenvalue with largest real part. When the largest eigenvalue of \(\mathbf{S}\) is negative, Eq.~\eqref{eq:etS} vanishes. In contrast, if the largest eigenvalue is zero, \(e^{t \mathbf{S}} \to A_{\tilde{k}}\), which means that the propagation through hidden pathways does not leak all its probability and the system can get stuck in a hidden part, never performing a new visible transition.

\bibliography{refs}

\begin{thebibliography}{87}%
\makeatletter
\providecommand \@ifxundefined [1]{%
 \@ifx{#1\undefined}
}%
\providecommand \@ifnum [1]{%
 \ifnum #1\expandafter \@firstoftwo
 \else \expandafter \@secondoftwo
 \fi
}%
\providecommand \@ifx [1]{%
 \ifx #1\expandafter \@firstoftwo
 \else \expandafter \@secondoftwo
 \fi
}%
\providecommand \natexlab [1]{#1}%
\providecommand \enquote  [1]{``#1''}%
\providecommand \bibnamefont  [1]{#1}%
\providecommand \bibfnamefont [1]{#1}%
\providecommand \citenamefont [1]{#1}%
\providecommand \href@noop [0]{\@secondoftwo}%
\providecommand \href [0]{\begingroup \@sanitize@url \@href}%
\providecommand \@href[1]{\@@startlink{#1}\@@href}%
\providecommand \@@href[1]{\endgroup#1\@@endlink}%
\providecommand \@sanitize@url [0]{\catcode `\\12\catcode `\$12\catcode
  `\&12\catcode `\#12\catcode `\^12\catcode `\_12\catcode `\%12\relax}%
\providecommand \@@startlink[1]{}%
\providecommand \@@endlink[0]{}%
\providecommand \url  [0]{\begingroup\@sanitize@url \@url }%
\providecommand \@url [1]{\endgroup\@href {#1}{\urlprefix }}%
\providecommand \urlprefix  [0]{URL }%
\providecommand \Eprint [0]{\href }%
\providecommand \doibase [0]{http://dx.doi.org/}%
\providecommand \selectlanguage [0]{\@gobble}%
\providecommand \bibinfo  [0]{\@secondoftwo}%
\providecommand \bibfield  [0]{\@secondoftwo}%
\providecommand \translation [1]{[#1]}%
\providecommand \BibitemOpen [0]{}%
\providecommand \bibitemStop [0]{}%
\providecommand \bibitemNoStop [0]{.\EOS\space}%
\providecommand \EOS [0]{\spacefactor3000\relax}%
\providecommand \BibitemShut  [1]{\csname bibitem#1\endcsname}%
\let\auto@bib@innerbib\@empty
\bibitem [{\citenamefont {Fang}\ \emph {et~al.}(2019)\citenamefont {Fang},
  \citenamefont {Kruse}, \citenamefont {Lu},\ and\ \citenamefont
  {Wang}}]{fangNonequilibriumPhysicsBiology2019}%
  \BibitemOpen
  \bibfield  {author} {\bibinfo {author} {\bibfnamefont {Xiaona}\ \bibnamefont
  {Fang}}, \bibinfo {author} {\bibfnamefont {Karsten}\ \bibnamefont {Kruse}},
  \bibinfo {author} {\bibfnamefont {Ting}\ \bibnamefont {Lu}}, \ and\ \bibinfo
  {author} {\bibfnamefont {Jin}\ \bibnamefont {Wang}},\ }\bibfield  {title}
  {\enquote {\bibinfo {title} {Nonequilibrium physics in biology},}\ }\href
  {\doibase 10.1103/RevModPhys.91.045004} {\bibfield  {journal} {\bibinfo
  {journal} {Reviews of Modern Physics}\ }\textbf {\bibinfo {volume} {91}},\
  \bibinfo {pages} {045004} (\bibinfo {year} {2019})}\BibitemShut {NoStop}%
\bibitem [{\citenamefont {Gnesotto}\ \emph {et~al.}(2018)\citenamefont
  {Gnesotto}, \citenamefont {Mura}, \citenamefont {Gladrow},\ and\
  \citenamefont {Broedersz}}]{gnesottoBrokenDetailedBalance2018}%
  \BibitemOpen
  \bibfield  {author} {\bibinfo {author} {\bibfnamefont {F.~S.}\ \bibnamefont
  {Gnesotto}}, \bibinfo {author} {\bibfnamefont {F.}~\bibnamefont {Mura}},
  \bibinfo {author} {\bibfnamefont {J.}~\bibnamefont {Gladrow}}, \ and\
  \bibinfo {author} {\bibfnamefont {C.~P.}\ \bibnamefont {Broedersz}},\
  }\bibfield  {title} {\enquote {\bibinfo {title} {Broken detailed balance and
  non-equilibrium dynamics in living systems: A review},}\ }\href {\doibase
  10.1088/1361-6633/aab3ed} {\bibfield  {journal} {\bibinfo  {journal} {Reports
  on Progress in Physics}\ }\textbf {\bibinfo {volume} {81}},\ \bibinfo {pages}
  {066601} (\bibinfo {year} {2018})}\BibitemShut {NoStop}%
\bibitem [{\citenamefont {Zoller}\ \emph {et~al.}(2022)\citenamefont {Zoller},
  \citenamefont {Gregor},\ and\ \citenamefont {Tka{\v
  c}ik}}]{zollerEukaryoticGeneRegulation2022}%
  \BibitemOpen
  \bibfield  {author} {\bibinfo {author} {\bibfnamefont {Benjamin}\
  \bibnamefont {Zoller}}, \bibinfo {author} {\bibfnamefont {Thomas}\
  \bibnamefont {Gregor}}, \ and\ \bibinfo {author} {\bibfnamefont {Ga{\v
  s}per}\ \bibnamefont {Tka{\v c}ik}},\ }\bibfield  {title} {\enquote {\bibinfo
  {title} {Eukaryotic gene regulation at equilibrium, or non?}}\ }\href
  {\doibase 10.1016/j.coisb.2022.100435} {\bibfield  {journal} {\bibinfo
  {journal} {Current Opinion in Systems Biology}\ }\textbf {\bibinfo {volume}
  {31}},\ \bibinfo {pages} {100435} (\bibinfo {year} {2022})}\BibitemShut
  {NoStop}%
\bibitem [{\citenamefont {Hartich}\ \emph {et~al.}(2015)\citenamefont
  {Hartich}, \citenamefont {Barato},\ and\ \citenamefont
  {Seifert}}]{hartichNonequilibriumSensingIts2015}%
  \BibitemOpen
  \bibfield  {author} {\bibinfo {author} {\bibfnamefont {David}\ \bibnamefont
  {Hartich}}, \bibinfo {author} {\bibfnamefont {Andre~C.}\ \bibnamefont
  {Barato}}, \ and\ \bibinfo {author} {\bibfnamefont {Udo}\ \bibnamefont
  {Seifert}},\ }\bibfield  {title} {\enquote {\bibinfo {title} {Nonequilibrium
  sensing and its analogy to kinetic proofreading},}\ }\href {\doibase
  10.1088/1367-2630/17/5/055026} {\bibfield  {journal} {\bibinfo  {journal}
  {New Journal of Physics}\ }\textbf {\bibinfo {volume} {17}},\ \bibinfo
  {pages} {055026} (\bibinfo {year} {2015})}\BibitemShut {NoStop}%
\bibitem [{\citenamefont {Yang}\ \emph {et~al.}(2021)\citenamefont {Yang},
  \citenamefont {Heinemann}, \citenamefont {Howard}, \citenamefont {Huber},
  \citenamefont {{Iyer-Biswas}}, \citenamefont {Le~Treut}, \citenamefont
  {Lynch}, \citenamefont {Montooth}, \citenamefont {Needleman}, \citenamefont
  {Pigolotti}, \citenamefont {Rodenfels}, \citenamefont {Ronceray},
  \citenamefont {Shankar}, \citenamefont {Tavassoly}, \citenamefont
  {Thutupalli}, \citenamefont {Titov}, \citenamefont {Wang},\ and\
  \citenamefont {Foster}}]{yangPhysicalBioenergeticsEnergy2021}%
  \BibitemOpen
  \bibfield  {author} {\bibinfo {author} {\bibfnamefont {Xingbo}\ \bibnamefont
  {Yang}}, \bibinfo {author} {\bibfnamefont {Matthias}\ \bibnamefont
  {Heinemann}}, \bibinfo {author} {\bibfnamefont {Jonathon}\ \bibnamefont
  {Howard}}, \bibinfo {author} {\bibfnamefont {Greg}\ \bibnamefont {Huber}},
  \bibinfo {author} {\bibfnamefont {Srividya}\ \bibnamefont {{Iyer-Biswas}}},
  \bibinfo {author} {\bibfnamefont {Guillaume}\ \bibnamefont {Le~Treut}},
  \bibinfo {author} {\bibfnamefont {Michael}\ \bibnamefont {Lynch}}, \bibinfo
  {author} {\bibfnamefont {Kristi~L.}\ \bibnamefont {Montooth}}, \bibinfo
  {author} {\bibfnamefont {Daniel~J.}\ \bibnamefont {Needleman}}, \bibinfo
  {author} {\bibfnamefont {Simone}\ \bibnamefont {Pigolotti}}, \bibinfo
  {author} {\bibfnamefont {Jonathan}\ \bibnamefont {Rodenfels}}, \bibinfo
  {author} {\bibfnamefont {Pierre}\ \bibnamefont {Ronceray}}, \bibinfo {author}
  {\bibfnamefont {Sadasivan}\ \bibnamefont {Shankar}}, \bibinfo {author}
  {\bibfnamefont {Iman}\ \bibnamefont {Tavassoly}}, \bibinfo {author}
  {\bibfnamefont {Shashi}\ \bibnamefont {Thutupalli}}, \bibinfo {author}
  {\bibfnamefont {Denis~V.}\ \bibnamefont {Titov}}, \bibinfo {author}
  {\bibfnamefont {Jin}\ \bibnamefont {Wang}}, \ and\ \bibinfo {author}
  {\bibfnamefont {Peter~J.}\ \bibnamefont {Foster}},\ }\bibfield  {title}
  {\enquote {\bibinfo {title} {Physical bioenergetics: {{Energy}} fluxes,
  budgets, and constraints in cells},}\ }\href {\doibase
  10.1073/pnas.2026786118} {\bibfield  {journal} {\bibinfo  {journal}
  {Proceedings of the National Academy of Sciences}\ }\textbf {\bibinfo
  {volume} {118}},\ \bibinfo {pages} {e2026786118} (\bibinfo {year}
  {2021})}\BibitemShut {NoStop}%
\bibitem [{\citenamefont {Jarzynski}(1997)}]{PhysRevLett.78.2690}%
  \BibitemOpen
  \bibfield  {author} {\bibinfo {author} {\bibfnamefont {C.}~\bibnamefont
  {Jarzynski}},\ }\bibfield  {title} {\enquote {\bibinfo {title}
  {Nonequilibrium equality for free energy differences},}\ }\href {\doibase
  10.1103/PhysRevLett.78.2690} {\bibfield  {journal} {\bibinfo  {journal}
  {Phys. Rev. Lett.}\ }\textbf {\bibinfo {volume} {78}},\ \bibinfo {pages}
  {2690--2693} (\bibinfo {year} {1997})}\BibitemShut {NoStop}%
\bibitem [{\citenamefont {Crooks}(1999)}]{PhysRevE.60.2721}%
  \BibitemOpen
  \bibfield  {author} {\bibinfo {author} {\bibfnamefont {Gavin~E.}\
  \bibnamefont {Crooks}},\ }\bibfield  {title} {\enquote {\bibinfo {title}
  {Entropy production fluctuation theorem and the nonequilibrium work relation
  for free energy differences},}\ }\href {\doibase 10.1103/PhysRevE.60.2721}
  {\bibfield  {journal} {\bibinfo  {journal} {Phys. Rev. E}\ }\textbf {\bibinfo
  {volume} {60}},\ \bibinfo {pages} {2721--2726} (\bibinfo {year}
  {1999})}\BibitemShut {NoStop}%
\bibitem [{\citenamefont {Evans}\ \emph {et~al.}(1993)\citenamefont {Evans},
  \citenamefont {Cohen},\ and\ \citenamefont {Morriss}}]{PhysRevLett.71.2401}%
  \BibitemOpen
  \bibfield  {author} {\bibinfo {author} {\bibfnamefont {Denis~J.}\
  \bibnamefont {Evans}}, \bibinfo {author} {\bibfnamefont {E.~G.~D.}\
  \bibnamefont {Cohen}}, \ and\ \bibinfo {author} {\bibfnamefont {G.~P.}\
  \bibnamefont {Morriss}},\ }\bibfield  {title} {\enquote {\bibinfo {title}
  {Probability of second law violations in shearing steady states},}\ }\href
  {\doibase 10.1103/PhysRevLett.71.2401} {\bibfield  {journal} {\bibinfo
  {journal} {Phys. Rev. Lett.}\ }\textbf {\bibinfo {volume} {71}},\ \bibinfo
  {pages} {2401--2404} (\bibinfo {year} {1993})}\BibitemShut {NoStop}%
\bibitem [{\citenamefont {Barato}\ and\ \citenamefont
  {Seifert}(2015)}]{PhysRevLett.114.158101}%
  \BibitemOpen
  \bibfield  {author} {\bibinfo {author} {\bibfnamefont {Andre~C.}\
  \bibnamefont {Barato}}\ and\ \bibinfo {author} {\bibfnamefont {Udo}\
  \bibnamefont {Seifert}},\ }\bibfield  {title} {\enquote {\bibinfo {title}
  {Thermodynamic uncertainty relation for biomolecular processes},}\ }\href
  {\doibase 10.1103/PhysRevLett.114.158101} {\bibfield  {journal} {\bibinfo
  {journal} {Phys. Rev. Lett.}\ }\textbf {\bibinfo {volume} {114}},\ \bibinfo
  {pages} {158101} (\bibinfo {year} {2015})}\BibitemShut {NoStop}%
\bibitem [{\citenamefont {Horowitz}\ and\ \citenamefont
  {Gingrich}(2020)}]{horowitz2020thermodynamic}%
  \BibitemOpen
  \bibfield  {author} {\bibinfo {author} {\bibfnamefont {Jordan~M}\
  \bibnamefont {Horowitz}}\ and\ \bibinfo {author} {\bibfnamefont {Todd~R}\
  \bibnamefont {Gingrich}},\ }\bibfield  {title} {\enquote {\bibinfo {title}
  {Thermodynamic uncertainty relations constrain non-equilibrium
  fluctuations},}\ }\href {\doibase https://doi.org/10.1038/s41567-019-0702-6}
  {\bibfield  {journal} {\bibinfo  {journal} {Nature Physics}\ }\textbf
  {\bibinfo {volume} {16}},\ \bibinfo {pages} {15--20} (\bibinfo {year}
  {2020})}\BibitemShut {NoStop}%
\bibitem [{\citenamefont {Falasco}\ and\ \citenamefont
  {Esposito}(2020)}]{PhysRevLett.125.120604}%
  \BibitemOpen
  \bibfield  {author} {\bibinfo {author} {\bibfnamefont {Gianmaria}\
  \bibnamefont {Falasco}}\ and\ \bibinfo {author} {\bibfnamefont
  {Massimiliano}\ \bibnamefont {Esposito}},\ }\bibfield  {title} {\enquote
  {\bibinfo {title} {Dissipation-time uncertainty relation},}\ }\href {\doibase
  10.1103/PhysRevLett.125.120604} {\bibfield  {journal} {\bibinfo  {journal}
  {Phys. Rev. Lett.}\ }\textbf {\bibinfo {volume} {125}},\ \bibinfo {pages}
  {120604} (\bibinfo {year} {2020})}\BibitemShut {NoStop}%
\bibitem [{\citenamefont {Shiraishi}\ \emph {et~al.}(2018)\citenamefont
  {Shiraishi}, \citenamefont {Funo},\ and\ \citenamefont
  {Saito}}]{PhysRevLett.121.070601}%
  \BibitemOpen
  \bibfield  {author} {\bibinfo {author} {\bibfnamefont {Naoto}\ \bibnamefont
  {Shiraishi}}, \bibinfo {author} {\bibfnamefont {Ken}\ \bibnamefont {Funo}}, \
  and\ \bibinfo {author} {\bibfnamefont {Keiji}\ \bibnamefont {Saito}},\
  }\bibfield  {title} {\enquote {\bibinfo {title} {Speed limit for classical
  stochastic processes},}\ }\href {\doibase 10.1103/PhysRevLett.121.070601}
  {\bibfield  {journal} {\bibinfo  {journal} {Phys. Rev. Lett.}\ }\textbf
  {\bibinfo {volume} {121}},\ \bibinfo {pages} {070601} (\bibinfo {year}
  {2018})}\BibitemShut {NoStop}%
\bibitem [{\citenamefont {Owen}\ \emph {et~al.}(2020)\citenamefont {Owen},
  \citenamefont {Gingrich},\ and\ \citenamefont
  {Horowitz}}]{PhysRevX.10.011066}%
  \BibitemOpen
  \bibfield  {author} {\bibinfo {author} {\bibfnamefont {Jeremy~A.}\
  \bibnamefont {Owen}}, \bibinfo {author} {\bibfnamefont {Todd~R.}\
  \bibnamefont {Gingrich}}, \ and\ \bibinfo {author} {\bibfnamefont
  {Jordan~M.}\ \bibnamefont {Horowitz}},\ }\bibfield  {title} {\enquote
  {\bibinfo {title} {Universal thermodynamic bounds on nonequilibrium response
  with biochemical applications},}\ }\href {\doibase
  10.1103/PhysRevX.10.011066} {\bibfield  {journal} {\bibinfo  {journal} {Phys.
  Rev. X}\ }\textbf {\bibinfo {volume} {10}},\ \bibinfo {pages} {011066}
  (\bibinfo {year} {2020})}\BibitemShut {NoStop}%
\bibitem [{\citenamefont {Basu}\ and\ \citenamefont {Maes}(2015)}]{Basu_2015}%
  \BibitemOpen
  \bibfield  {author} {\bibinfo {author} {\bibfnamefont {Urna}\ \bibnamefont
  {Basu}}\ and\ \bibinfo {author} {\bibfnamefont {Christian}\ \bibnamefont
  {Maes}},\ }\bibfield  {title} {\enquote {\bibinfo {title} {Nonequilibrium
  response and frenesy},}\ }\href {\doibase 10.1088/1742-6596/638/1/012001}
  {\bibfield  {journal} {\bibinfo  {journal} {Journal of Physics: Conference
  Series}\ }\textbf {\bibinfo {volume} {638}},\ \bibinfo {pages} {012001}
  (\bibinfo {year} {2015})}\BibitemShut {NoStop}%
\bibitem [{\citenamefont {Kawaguchi}\ and\ \citenamefont
  {Nakayama}(2013)}]{PhysRevE.88.022147}%
  \BibitemOpen
  \bibfield  {author} {\bibinfo {author} {\bibfnamefont {Kyogo}\ \bibnamefont
  {Kawaguchi}}\ and\ \bibinfo {author} {\bibfnamefont {Yohei}\ \bibnamefont
  {Nakayama}},\ }\bibfield  {title} {\enquote {\bibinfo {title} {Fluctuation
  theorem for hidden entropy production},}\ }\href {\doibase
  10.1103/PhysRevE.88.022147} {\bibfield  {journal} {\bibinfo  {journal} {Phys.
  Rev. E}\ }\textbf {\bibinfo {volume} {88}},\ \bibinfo {pages} {022147}
  (\bibinfo {year} {2013})}\BibitemShut {NoStop}%
\bibitem [{\citenamefont {Herpich}\ \emph {et~al.}(2020)\citenamefont
  {Herpich}, \citenamefont {Shayanfard},\ and\ \citenamefont
  {Esposito}}]{herpichEffectiveThermodynamicsTwo2020}%
  \BibitemOpen
  \bibfield  {author} {\bibinfo {author} {\bibfnamefont {Tim}\ \bibnamefont
  {Herpich}}, \bibinfo {author} {\bibfnamefont {Kamran}\ \bibnamefont
  {Shayanfard}}, \ and\ \bibinfo {author} {\bibfnamefont {Massimiliano}\
  \bibnamefont {Esposito}},\ }\bibfield  {title} {\enquote {\bibinfo {title}
  {Effective thermodynamics of two interacting underdamped {{Brownian}}
  particles},}\ }\href {\doibase 10.1103/PhysRevE.101.022116} {\bibfield
  {journal} {\bibinfo  {journal} {Physical Review E}\ }\textbf {\bibinfo
  {volume} {101}},\ \bibinfo {pages} {022116} (\bibinfo {year}
  {2020})}\BibitemShut {NoStop}%
\bibitem [{\citenamefont {Manikandan}\ \emph {et~al.}(2020)\citenamefont
  {Manikandan}, \citenamefont {Gupta},\ and\ \citenamefont
  {Krishnamurthy}}]{PhysRevLett.124.120603}%
  \BibitemOpen
  \bibfield  {author} {\bibinfo {author} {\bibfnamefont {Sreekanth~K.}\
  \bibnamefont {Manikandan}}, \bibinfo {author} {\bibfnamefont {Deepak}\
  \bibnamefont {Gupta}}, \ and\ \bibinfo {author} {\bibfnamefont {Supriya}\
  \bibnamefont {Krishnamurthy}},\ }\bibfield  {title} {\enquote {\bibinfo
  {title} {Inferring entropy production from short experiments},}\ }\href
  {\doibase 10.1103/PhysRevLett.124.120603} {\bibfield  {journal} {\bibinfo
  {journal} {Phys. Rev. Lett.}\ }\textbf {\bibinfo {volume} {124}},\ \bibinfo
  {pages} {120603} (\bibinfo {year} {2020})}\BibitemShut {NoStop}%
\bibitem [{\citenamefont {Busiello}\ and\ \citenamefont
  {Pigolotti}(2019)}]{PhysRevE.100.060102}%
  \BibitemOpen
  \bibfield  {author} {\bibinfo {author} {\bibfnamefont {Daniel~Maria}\
  \bibnamefont {Busiello}}\ and\ \bibinfo {author} {\bibfnamefont {Simone}\
  \bibnamefont {Pigolotti}},\ }\bibfield  {title} {\enquote {\bibinfo {title}
  {Hyperaccurate currents in stochastic thermodynamics},}\ }\href {\doibase
  10.1103/PhysRevE.100.060102} {\bibfield  {journal} {\bibinfo  {journal}
  {Phys. Rev. E}\ }\textbf {\bibinfo {volume} {100}},\ \bibinfo {pages}
  {060102} (\bibinfo {year} {2019})}\BibitemShut {NoStop}%
\bibitem [{\citenamefont {Li}\ \emph {et~al.}(2019)\citenamefont {Li},
  \citenamefont {Horowitz}, \citenamefont {Gingrich},\ and\ \citenamefont
  {Fakhri}}]{li2019quantifying}%
  \BibitemOpen
  \bibfield  {author} {\bibinfo {author} {\bibfnamefont {Junang}\ \bibnamefont
  {Li}}, \bibinfo {author} {\bibfnamefont {Jordan~M}\ \bibnamefont {Horowitz}},
  \bibinfo {author} {\bibfnamefont {Todd~R}\ \bibnamefont {Gingrich}}, \ and\
  \bibinfo {author} {\bibfnamefont {Nikta}\ \bibnamefont {Fakhri}},\ }\bibfield
   {title} {\enquote {\bibinfo {title} {Quantifying dissipation using
  fluctuating currents},}\ }\href {\doibase
  https://doi.org/10.1038/s41467-019-09631-x} {\bibfield  {journal} {\bibinfo
  {journal} {Nature communications}\ }\textbf {\bibinfo {volume} {10}},\
  \bibinfo {pages} {1666} (\bibinfo {year} {2019})}\BibitemShut {NoStop}%
\bibitem [{\citenamefont {Van~Vu}\ \emph {et~al.}(2020)\citenamefont {Van~Vu},
  \citenamefont {Vo},\ and\ \citenamefont {Hasegawa}}]{PhysRevE.101.042138}%
  \BibitemOpen
  \bibfield  {author} {\bibinfo {author} {\bibfnamefont {Tan}\ \bibnamefont
  {Van~Vu}}, \bibinfo {author} {\bibfnamefont {Van~Tuan}\ \bibnamefont {Vo}}, \
  and\ \bibinfo {author} {\bibfnamefont {Yoshihiko}\ \bibnamefont {Hasegawa}},\
  }\bibfield  {title} {\enquote {\bibinfo {title} {Entropy production
  estimation with optimal current},}\ }\href {\doibase
  10.1103/PhysRevE.101.042138} {\bibfield  {journal} {\bibinfo  {journal}
  {Phys. Rev. E}\ }\textbf {\bibinfo {volume} {101}},\ \bibinfo {pages}
  {042138} (\bibinfo {year} {2020})}\BibitemShut {NoStop}%
\bibitem [{\citenamefont {Skinner}\ and\ \citenamefont
  {Dunkel}(2021)}]{skinner2021improved}%
  \BibitemOpen
  \bibfield  {author} {\bibinfo {author} {\bibfnamefont {Dominic~J}\
  \bibnamefont {Skinner}}\ and\ \bibinfo {author} {\bibfnamefont {J{\"o}rn}\
  \bibnamefont {Dunkel}},\ }\bibfield  {title} {\enquote {\bibinfo {title}
  {Improved bounds on entropy production in living systems},}\ }\href {\doibase
  https://doi.org/10.1073/pnas.2024300118} {\bibfield  {journal} {\bibinfo
  {journal} {Proceedings of the National Academy of Sciences}\ }\textbf
  {\bibinfo {volume} {118}},\ \bibinfo {pages} {e2024300118} (\bibinfo {year}
  {2021})}\BibitemShut {NoStop}%
\bibitem [{\citenamefont {Shiraishi}\ and\ \citenamefont
  {Sagawa}(2015)}]{PhysRevE.91.012130}%
  \BibitemOpen
  \bibfield  {author} {\bibinfo {author} {\bibfnamefont {Naoto}\ \bibnamefont
  {Shiraishi}}\ and\ \bibinfo {author} {\bibfnamefont {Takahiro}\ \bibnamefont
  {Sagawa}},\ }\bibfield  {title} {\enquote {\bibinfo {title} {Fluctuation
  theorem for partially masked nonequilibrium dynamics},}\ }\href {\doibase
  10.1103/PhysRevE.91.012130} {\bibfield  {journal} {\bibinfo  {journal} {Phys.
  Rev. E}\ }\textbf {\bibinfo {volume} {91}},\ \bibinfo {pages} {012130}
  (\bibinfo {year} {2015})}\BibitemShut {NoStop}%
\bibitem [{\citenamefont {Bisker}\ \emph {et~al.}(2017)\citenamefont {Bisker},
  \citenamefont {Polettini}, \citenamefont {Gingrich},\ and\ \citenamefont
  {Horowitz}}]{Bisker_2017}%
  \BibitemOpen
  \bibfield  {author} {\bibinfo {author} {\bibfnamefont {Gili}\ \bibnamefont
  {Bisker}}, \bibinfo {author} {\bibfnamefont {Matteo}\ \bibnamefont
  {Polettini}}, \bibinfo {author} {\bibfnamefont {Todd~R}\ \bibnamefont
  {Gingrich}}, \ and\ \bibinfo {author} {\bibfnamefont {Jordan~M}\ \bibnamefont
  {Horowitz}},\ }\bibfield  {title} {\enquote {\bibinfo {title} {Hierarchical
  bounds on entropy production inferred from partial information},}\ }\href
  {\doibase 10.1088/1742-5468/aa8c0d} {\bibfield  {journal} {\bibinfo
  {journal} {Journal of Statistical Mechanics: Theory and Experiment}\ }\textbf
  {\bibinfo {volume} {2017}},\ \bibinfo {pages} {093210} (\bibinfo {year}
  {2017})}\BibitemShut {NoStop}%
\bibitem [{\citenamefont {Ehrich}(2021)}]{Ehrich_2021}%
  \BibitemOpen
  \bibfield  {author} {\bibinfo {author} {\bibfnamefont {Jannik}\ \bibnamefont
  {Ehrich}},\ }\bibfield  {title} {\enquote {\bibinfo {title} {Tightest bound
  on hidden entropy production from partially observed dynamics},}\ }\href
  {\doibase 10.1088/1742-5468/ac150e} {\bibfield  {journal} {\bibinfo
  {journal} {Journal of Statistical Mechanics: Theory and Experiment}\ }\textbf
  {\bibinfo {volume} {2021}},\ \bibinfo {pages} {083214} (\bibinfo {year}
  {2021})}\BibitemShut {NoStop}%
\bibitem [{\citenamefont {Blom}\ \emph {et~al.}(2024)\citenamefont {Blom},
  \citenamefont {Song}, \citenamefont {Vouga}, \citenamefont {Godec},\ and\
  \citenamefont {Makarov}}]{blom2023milestoning}%
  \BibitemOpen
  \bibfield  {author} {\bibinfo {author} {\bibfnamefont {Kristian}\
  \bibnamefont {Blom}}, \bibinfo {author} {\bibfnamefont {Kevin}\ \bibnamefont
  {Song}}, \bibinfo {author} {\bibfnamefont {Etienne}\ \bibnamefont {Vouga}},
  \bibinfo {author} {\bibfnamefont {Alja{\v z}}\ \bibnamefont {Godec}}, \ and\
  \bibinfo {author} {\bibfnamefont {Dmitrii~E.}\ \bibnamefont {Makarov}},\
  }\bibfield  {title} {\enquote {\bibinfo {title} {Milestoning estimators of
  dissipation in systems observed at a coarse resolution},}\ }\href {\doibase
  10.1073/pnas.2318333121} {\bibfield  {journal} {\bibinfo  {journal}
  {Proceedings of the National Academy of Sciences}\ }\textbf {\bibinfo
  {volume} {121}},\ \bibinfo {pages} {e2318333121} (\bibinfo {year}
  {2024})}\BibitemShut {NoStop}%
\bibitem [{\citenamefont {Rold\'an}\ and\ \citenamefont
  {Parrondo}(2010)}]{PhysRevLett.105.150607}%
  \BibitemOpen
  \bibfield  {author} {\bibinfo {author} {\bibfnamefont {\'Edgar}\ \bibnamefont
  {Rold\'an}}\ and\ \bibinfo {author} {\bibfnamefont {Juan M.~R.}\ \bibnamefont
  {Parrondo}},\ }\bibfield  {title} {\enquote {\bibinfo {title} {Estimating
  dissipation from single stationary trajectories},}\ }\href {\doibase
  10.1103/PhysRevLett.105.150607} {\bibfield  {journal} {\bibinfo  {journal}
  {Phys. Rev. Lett.}\ }\textbf {\bibinfo {volume} {105}},\ \bibinfo {pages}
  {150607} (\bibinfo {year} {2010})}\BibitemShut {NoStop}%
\bibitem [{\citenamefont {Mart{\'\i}nez}\ \emph {et~al.}(2019)\citenamefont
  {Mart{\'\i}nez}, \citenamefont {Bisker}, \citenamefont {Horowitz},\ and\
  \citenamefont {Parrondo}}]{martinez2019inferring}%
  \BibitemOpen
  \bibfield  {author} {\bibinfo {author} {\bibfnamefont {Ignacio~A}\
  \bibnamefont {Mart{\'\i}nez}}, \bibinfo {author} {\bibfnamefont {Gili}\
  \bibnamefont {Bisker}}, \bibinfo {author} {\bibfnamefont {Jordan~M}\
  \bibnamefont {Horowitz}}, \ and\ \bibinfo {author} {\bibfnamefont {Juan~MR}\
  \bibnamefont {Parrondo}},\ }\bibfield  {title} {\enquote {\bibinfo {title}
  {Inferring broken detailed balance in the absence of observable currents},}\
  }\href {\doibase https://doi.org/10.1038/s41467-019-11051-w} {\bibfield
  {journal} {\bibinfo  {journal} {Nature communications}\ }\textbf {\bibinfo
  {volume} {10}},\ \bibinfo {pages} {3542} (\bibinfo {year}
  {2019})}\BibitemShut {NoStop}%
\bibitem [{\citenamefont {Esposito}(2012)}]{PhysRevE.85.041125}%
  \BibitemOpen
  \bibfield  {author} {\bibinfo {author} {\bibfnamefont {Massimiliano}\
  \bibnamefont {Esposito}},\ }\bibfield  {title} {\enquote {\bibinfo {title}
  {Stochastic thermodynamics under coarse graining},}\ }\href {\doibase
  10.1103/PhysRevE.85.041125} {\bibfield  {journal} {\bibinfo  {journal} {Phys.
  Rev. E}\ }\textbf {\bibinfo {volume} {85}},\ \bibinfo {pages} {041125}
  (\bibinfo {year} {2012})}\BibitemShut {NoStop}%
\bibitem [{\citenamefont {Bo}\ and\ \citenamefont
  {Celani}(2014)}]{boEntropyProductionStochastic2014}%
  \BibitemOpen
  \bibfield  {author} {\bibinfo {author} {\bibfnamefont {Stefano}\ \bibnamefont
  {Bo}}\ and\ \bibinfo {author} {\bibfnamefont {Antonio}\ \bibnamefont
  {Celani}},\ }\bibfield  {title} {\enquote {\bibinfo {title} {Entropy
  {{Production}} in {{Stochastic Systems}} with {{Fast}} and {{Slow
  Time-Scales}}},}\ }\href {\doibase 10.1007/s10955-014-0922-1} {\bibfield
  {journal} {\bibinfo  {journal} {Journal of Statistical Physics}\ }\textbf
  {\bibinfo {volume} {154}},\ \bibinfo {pages} {1325--1351} (\bibinfo {year}
  {2014})}\BibitemShut {NoStop}%
\bibitem [{\citenamefont {Rahav}\ and\ \citenamefont
  {Jarzynski}(2007)}]{rahavFluctuationRelationsCoarsegraining2007}%
  \BibitemOpen
  \bibfield  {author} {\bibinfo {author} {\bibfnamefont {Saar}\ \bibnamefont
  {Rahav}}\ and\ \bibinfo {author} {\bibfnamefont {Christopher}\ \bibnamefont
  {Jarzynski}},\ }\bibfield  {title} {\enquote {\bibinfo {title} {Fluctuation
  relations and coarse-graining},}\ }\href {\doibase
  10.1088/1742-5468/2007/09/P09012} {\bibfield  {journal} {\bibinfo  {journal}
  {Journal of Statistical Mechanics: Theory and Experiment}\ }\textbf {\bibinfo
  {volume} {2007}},\ \bibinfo {pages} {P09012} (\bibinfo {year}
  {2007})}\BibitemShut {NoStop}%
\bibitem [{\citenamefont {Harunari}\ \emph {et~al.}(2022)\citenamefont
  {Harunari}, \citenamefont {Dutta}, \citenamefont {Polettini},\ and\
  \citenamefont {Rold\'an}}]{PhysRevX.12.041026}%
  \BibitemOpen
  \bibfield  {author} {\bibinfo {author} {\bibfnamefont {Pedro~E.}\
  \bibnamefont {Harunari}}, \bibinfo {author} {\bibfnamefont {Annwesha}\
  \bibnamefont {Dutta}}, \bibinfo {author} {\bibfnamefont {Matteo}\
  \bibnamefont {Polettini}}, \ and\ \bibinfo {author} {\bibfnamefont {\'Edgar}\
  \bibnamefont {Rold\'an}},\ }\bibfield  {title} {\enquote {\bibinfo {title}
  {What to learn from a few visible transitions' statistics?}}\ }\href
  {\doibase 10.1103/PhysRevX.12.041026} {\bibfield  {journal} {\bibinfo
  {journal} {Phys. Rev. X}\ }\textbf {\bibinfo {volume} {12}},\ \bibinfo
  {pages} {041026} (\bibinfo {year} {2022})}\BibitemShut {NoStop}%
\bibitem [{\citenamefont {van~der Meer}\ \emph {et~al.}(2022)\citenamefont
  {van~der Meer}, \citenamefont {Ertel},\ and\ \citenamefont
  {Seifert}}]{PhysRevX.12.031025}%
  \BibitemOpen
  \bibfield  {author} {\bibinfo {author} {\bibfnamefont {Jann}\ \bibnamefont
  {van~der Meer}}, \bibinfo {author} {\bibfnamefont {Benjamin}\ \bibnamefont
  {Ertel}}, \ and\ \bibinfo {author} {\bibfnamefont {Udo}\ \bibnamefont
  {Seifert}},\ }\bibfield  {title} {\enquote {\bibinfo {title} {Thermodynamic
  inference in partially accessible markov networks: A unifying perspective
  from transition-based waiting time distributions},}\ }\href {\doibase
  10.1103/PhysRevX.12.031025} {\bibfield  {journal} {\bibinfo  {journal} {Phys.
  Rev. X}\ }\textbf {\bibinfo {volume} {12}},\ \bibinfo {pages} {031025}
  (\bibinfo {year} {2022})}\BibitemShut {NoStop}%
\bibitem [{\citenamefont {van~der Meer}\ \emph {et~al.}(2023)\citenamefont
  {van~der Meer}, \citenamefont {Deg\"unther},\ and\ \citenamefont
  {Seifert}}]{PhysRevLett.130.257101}%
  \BibitemOpen
  \bibfield  {author} {\bibinfo {author} {\bibfnamefont {Jann}\ \bibnamefont
  {van~der Meer}}, \bibinfo {author} {\bibfnamefont {Julius}\ \bibnamefont
  {Deg\"unther}}, \ and\ \bibinfo {author} {\bibfnamefont {Udo}\ \bibnamefont
  {Seifert}},\ }\bibfield  {title} {\enquote {\bibinfo {title} {Time-resolved
  statistics of snippets as general framework for model-free entropy
  estimators},}\ }\href {\doibase 10.1103/PhysRevLett.130.257101} {\bibfield
  {journal} {\bibinfo  {journal} {Phys. Rev. Lett.}\ }\textbf {\bibinfo
  {volume} {130}},\ \bibinfo {pages} {257101} (\bibinfo {year}
  {2023})}\BibitemShut {NoStop}%
\bibitem [{\citenamefont {Harunari}\ \emph {et~al.}(2023)\citenamefont
  {Harunari}, \citenamefont {Garilli},\ and\ \citenamefont
  {Polettini}}]{PhysRevE.107.L042105}%
  \BibitemOpen
  \bibfield  {author} {\bibinfo {author} {\bibfnamefont {Pedro~E.}\
  \bibnamefont {Harunari}}, \bibinfo {author} {\bibfnamefont {Alberto}\
  \bibnamefont {Garilli}}, \ and\ \bibinfo {author} {\bibfnamefont {Matteo}\
  \bibnamefont {Polettini}},\ }\bibfield  {title} {\enquote {\bibinfo {title}
  {Beat of a current},}\ }\href {\doibase 10.1103/PhysRevE.107.L042105}
  {\bibfield  {journal} {\bibinfo  {journal} {Phys. Rev. E}\ }\textbf {\bibinfo
  {volume} {107}},\ \bibinfo {pages} {L042105} (\bibinfo {year}
  {2023})}\BibitemShut {NoStop}%
\bibitem [{\citenamefont {Garilli}\ \emph {et~al.}(2023)\citenamefont
  {Garilli}, \citenamefont {Harunari},\ and\ \citenamefont
  {Polettini}}]{garilli2023fluctuation}%
  \BibitemOpen
  \bibfield  {author} {\bibinfo {author} {\bibfnamefont {Alberto}\ \bibnamefont
  {Garilli}}, \bibinfo {author} {\bibfnamefont {Pedro~E.}\ \bibnamefont
  {Harunari}}, \ and\ \bibinfo {author} {\bibfnamefont {Matteo}\ \bibnamefont
  {Polettini}},\ }\href@noop {} {\enquote {\bibinfo {title} {Fluctuation
  relations for a few observable currents at their own beat},}\ } (\bibinfo
  {year} {2023}),\ \Eprint {http://arxiv.org/abs/2312.07505} {arXiv:2312.07505
  [cond-mat.stat-mech]} \BibitemShut {NoStop}%
\bibitem [{\citenamefont {Pietzonka}\ and\ \citenamefont
  {Coghi}(2024)}]{pietzonkaThermodynamicCostPrecision2023}%
  \BibitemOpen
  \bibfield  {author} {\bibinfo {author} {\bibfnamefont {Patrick}\ \bibnamefont
  {Pietzonka}}\ and\ \bibinfo {author} {\bibfnamefont {Francesco}\ \bibnamefont
  {Coghi}},\ }\bibfield  {title} {\enquote {\bibinfo {title} {Thermodynamic
  cost for precision of general counting observables},}\ }\href {\doibase
  10.1103/PhysRevE.109.064128} {\bibfield  {journal} {\bibinfo  {journal}
  {Physical Review E}\ }\textbf {\bibinfo {volume} {109}},\ \bibinfo {pages}
  {064128} (\bibinfo {year} {2024})}\BibitemShut {NoStop}%
\bibitem [{\citenamefont {Maes}(2017)}]{maesFreneticBoundsEntropy2017}%
  \BibitemOpen
  \bibfield  {author} {\bibinfo {author} {\bibfnamefont {Christian}\
  \bibnamefont {Maes}},\ }\bibfield  {title} {\enquote {\bibinfo {title}
  {Frenetic {{Bounds}} on the {{Entropy Production}}},}\ }\href {\doibase
  10.1103/PhysRevLett.119.160601} {\bibfield  {journal} {\bibinfo  {journal}
  {Physical Review Letters}\ }\textbf {\bibinfo {volume} {119}},\ \bibinfo
  {pages} {160601} (\bibinfo {year} {2017})}\BibitemShut {NoStop}%
\bibitem [{\citenamefont {Berezhkovskii}\ and\ \citenamefont
  {Makarov}(2019)}]{berezhkovskii_forwardbackward_2019}%
  \BibitemOpen
  \bibfield  {author} {\bibinfo {author} {\bibfnamefont {Alexander~M.}\
  \bibnamefont {Berezhkovskii}}\ and\ \bibinfo {author} {\bibfnamefont
  {Dmitrii~E.}\ \bibnamefont {Makarov}},\ }\bibfield  {title} {\enquote
  {\bibinfo {title} {On the forward/backward symmetry of transition path time
  distributions in nonequilibrium systems},}\ }\href {\doibase
  10.1063/1.5109293} {\bibfield  {journal} {\bibinfo  {journal} {The Journal of
  Chemical Physics}\ }\textbf {\bibinfo {volume} {151}},\ \bibinfo {pages}
  {065102} (\bibinfo {year} {2019})}\BibitemShut {NoStop}%
\bibitem [{\citenamefont {Cristadoro}\ \emph {et~al.}(2023)\citenamefont
  {Cristadoro}, \citenamefont {Degli~Esposti}, \citenamefont {Jak{\v
  s}i{\'c}},\ and\ \citenamefont
  {Raqu{\'e}pas}}]{cristadoroRecurrenceTimesWaiting2023}%
  \BibitemOpen
  \bibfield  {author} {\bibinfo {author} {\bibfnamefont {Giampaolo}\
  \bibnamefont {Cristadoro}}, \bibinfo {author} {\bibfnamefont {Mirko}\
  \bibnamefont {Degli~Esposti}}, \bibinfo {author} {\bibfnamefont {Vojkan}\
  \bibnamefont {Jak{\v s}i{\'c}}}, \ and\ \bibinfo {author} {\bibfnamefont
  {Renaud}\ \bibnamefont {Raqu{\'e}pas}},\ }\bibfield  {title} {\enquote
  {\bibinfo {title} {Recurrence times, waiting times and universal entropy
  production estimators},}\ }\href {\doibase 10.1007/s11005-023-01640-8}
  {\bibfield  {journal} {\bibinfo  {journal} {Letters in Mathematical Physics}\
  }\textbf {\bibinfo {volume} {113}},\ \bibinfo {pages} {19} (\bibinfo {year}
  {2023})}\BibitemShut {NoStop}%
\bibitem [{\citenamefont {Pagare}\ \emph {et~al.}(2024)\citenamefont {Pagare},
  \citenamefont {Zhang}, \citenamefont {Zheng},\ and\ \citenamefont
  {Lu}}]{pagareStochasticDistinguishabilityMarkovian2024a}%
  \BibitemOpen
  \bibfield  {author} {\bibinfo {author} {\bibfnamefont {Asawari}\ \bibnamefont
  {Pagare}}, \bibinfo {author} {\bibfnamefont {Zhongmin}\ \bibnamefont
  {Zhang}}, \bibinfo {author} {\bibfnamefont {Jiming}\ \bibnamefont {Zheng}}, \
  and\ \bibinfo {author} {\bibfnamefont {Zhiyue}\ \bibnamefont {Lu}},\
  }\bibfield  {title} {\enquote {\bibinfo {title} {Stochastic
  distinguishability of {{Markovian}} trajectories},}\ }\href {\doibase
  10.1063/5.0203335} {\bibfield  {journal} {\bibinfo  {journal} {The Journal of
  Chemical Physics}\ }\textbf {\bibinfo {volume} {160}},\ \bibinfo {pages}
  {171101} (\bibinfo {year} {2024})}\BibitemShut {NoStop}%
\bibitem [{\citenamefont {Singh}\ and\ \citenamefont
  {Proesmans}(2023)}]{singh2023inferring}%
  \BibitemOpen
  \bibfield  {author} {\bibinfo {author} {\bibfnamefont {Prashant}\
  \bibnamefont {Singh}}\ and\ \bibinfo {author} {\bibfnamefont {Karel}\
  \bibnamefont {Proesmans}},\ }\href@noop {} {\enquote {\bibinfo {title}
  {Inferring entropy production from time-dependent moments},}\ } (\bibinfo
  {year} {2023}),\ \Eprint {http://arxiv.org/abs/2310.16627} {arXiv:2310.16627
  [cond-mat.stat-mech]} \BibitemShut {NoStop}%
\bibitem [{\citenamefont {Ferri-Cortés}\ \emph {et~al.}(2023)\citenamefont
  {Ferri-Cortés}, \citenamefont {Almanza-Marrero}, \citenamefont {López},
  \citenamefont {Zambrini},\ and\ \citenamefont
  {Manzano}}]{ferricortés2023entropy}%
  \BibitemOpen
  \bibfield  {author} {\bibinfo {author} {\bibfnamefont {Mar}\ \bibnamefont
  {Ferri-Cortés}}, \bibinfo {author} {\bibfnamefont {Jose~A.}\ \bibnamefont
  {Almanza-Marrero}}, \bibinfo {author} {\bibfnamefont {Rosa}\ \bibnamefont
  {López}}, \bibinfo {author} {\bibfnamefont {Roberta}\ \bibnamefont
  {Zambrini}}, \ and\ \bibinfo {author} {\bibfnamefont {Gonzalo}\ \bibnamefont
  {Manzano}},\ }\href@noop {} {\enquote {\bibinfo {title} {Entropy production
  and fluctuation theorems for monitored quantum systems under imperfect
  detection},}\ } (\bibinfo {year} {2023}),\ \Eprint
  {http://arxiv.org/abs/2308.08491} {arXiv:2308.08491 [quant-ph]} \BibitemShut
  {NoStop}%
\bibitem [{\citenamefont {Liang}\ and\ \citenamefont
  {Pigolotti}(2023)}]{liang_thermodynamic_2023}%
  \BibitemOpen
  \bibfield  {author} {\bibinfo {author} {\bibfnamefont {Shiling}\ \bibnamefont
  {Liang}}\ and\ \bibinfo {author} {\bibfnamefont {Simone}\ \bibnamefont
  {Pigolotti}},\ }\bibfield  {title} {\enquote {\bibinfo {title} {Thermodynamic
  bounds on time-reversal asymmetry},}\ }\href {\doibase
  10.1103/PhysRevE.108.L062101} {\bibfield  {journal} {\bibinfo  {journal}
  {Physical Review E}\ }\textbf {\bibinfo {volume} {108}},\ \bibinfo {pages}
  {L062101} (\bibinfo {year} {2023})}\BibitemShut {NoStop}%
\bibitem [{\citenamefont {Ertel}\ and\ \citenamefont
  {Seifert}(2024)}]{ertel2023estimator}%
  \BibitemOpen
  \bibfield  {author} {\bibinfo {author} {\bibfnamefont {Benjamin}\
  \bibnamefont {Ertel}}\ and\ \bibinfo {author} {\bibfnamefont {Udo}\
  \bibnamefont {Seifert}},\ }\bibfield  {title} {\enquote {\bibinfo {title}
  {Estimator of entropy production for partially accessible {{Markov}} networks
  based on the observation of blurred transitions},}\ }\href {\doibase
  10.1103/PhysRevE.109.054109} {\bibfield  {journal} {\bibinfo  {journal}
  {Physical Review E}\ }\textbf {\bibinfo {volume} {109}},\ \bibinfo {pages}
  {054109} (\bibinfo {year} {2024})}\BibitemShut {NoStop}%
\bibitem [{\citenamefont {Deg{\"u}nther}\ \emph {et~al.}(2024)\citenamefont
  {Deg{\"u}nther}, \citenamefont {Van Der~Meer},\ and\ \citenamefont
  {Seifert}}]{degunther2023fluctuating}%
  \BibitemOpen
  \bibfield  {author} {\bibinfo {author} {\bibfnamefont {Julius}\ \bibnamefont
  {Deg{\"u}nther}}, \bibinfo {author} {\bibfnamefont {Jann}\ \bibnamefont {Van
  Der~Meer}}, \ and\ \bibinfo {author} {\bibfnamefont {Udo}\ \bibnamefont
  {Seifert}},\ }\bibfield  {title} {\enquote {\bibinfo {title} {Fluctuating
  entropy production on the coarse-grained level: {{Inference}} and
  localization of irreversibility},}\ }\href {\doibase
  10.1103/PhysRevResearch.6.023175} {\bibfield  {journal} {\bibinfo  {journal}
  {Physical Review Research}\ }\textbf {\bibinfo {volume} {6}},\ \bibinfo
  {pages} {023175} (\bibinfo {year} {2024})}\BibitemShut {NoStop}%
\bibitem [{\citenamefont {Yu}\ \emph {et~al.}(2021)\citenamefont {Yu},
  \citenamefont {Zhang},\ and\ \citenamefont {Tu}}]{yuInversePowerLaw2021}%
  \BibitemOpen
  \bibfield  {author} {\bibinfo {author} {\bibfnamefont {Qiwei}\ \bibnamefont
  {Yu}}, \bibinfo {author} {\bibfnamefont {Dongliang}\ \bibnamefont {Zhang}}, \
  and\ \bibinfo {author} {\bibfnamefont {Yuhai}\ \bibnamefont {Tu}},\
  }\bibfield  {title} {\enquote {\bibinfo {title} {Inverse {{Power Law
  Scaling}} of {{Energy Dissipation Rate}} in {{Nonequilibrium Reaction
  Networks}}},}\ }\href {\doibase 10.1103/PhysRevLett.126.080601} {\bibfield
  {journal} {\bibinfo  {journal} {Physical Review Letters}\ }\textbf {\bibinfo
  {volume} {126}},\ \bibinfo {pages} {080601} (\bibinfo {year}
  {2021})}\BibitemShut {NoStop}%
\bibitem [{\citenamefont {Baiesi}\ \emph {et~al.}(2023)\citenamefont {Baiesi},
  \citenamefont {Falasco},\ and\ \citenamefont
  {Nishiyama}}]{baiesiEffectiveEstimationEntropy2023}%
  \BibitemOpen
  \bibfield  {author} {\bibinfo {author} {\bibfnamefont {Marco}\ \bibnamefont
  {Baiesi}}, \bibinfo {author} {\bibfnamefont {Gianmaria}\ \bibnamefont
  {Falasco}}, \ and\ \bibinfo {author} {\bibfnamefont {Tomohiro}\ \bibnamefont
  {Nishiyama}},\ }\href {\doibase 10.48550/arXiv.2305.04657} {\enquote
  {\bibinfo {title} {Effective estimation of entropy production with lacking
  data},}\ } (\bibinfo {year} {2023}),\ \Eprint
  {http://arxiv.org/abs/2305.04657} {arxiv:2305.04657 [cond-mat]} \BibitemShut
  {NoStop}%
\bibitem [{\citenamefont {Hartich}\ and\ \citenamefont
  {Godec}(2021)}]{hartichEmergentMemoryKinetic2021}%
  \BibitemOpen
  \bibfield  {author} {\bibinfo {author} {\bibfnamefont {David}\ \bibnamefont
  {Hartich}}\ and\ \bibinfo {author} {\bibfnamefont {Alja{\v z}}\ \bibnamefont
  {Godec}},\ }\bibfield  {title} {\enquote {\bibinfo {title} {Emergent
  {{Memory}} and {{Kinetic Hysteresis}} in {{Strongly Driven Networks}}},}\
  }\href {\doibase 10.1103/PhysRevX.11.041047} {\bibfield  {journal} {\bibinfo
  {journal} {Physical Review X}\ }\textbf {\bibinfo {volume} {11}},\ \bibinfo
  {pages} {041047} (\bibinfo {year} {2021})}\BibitemShut {NoStop}%
\bibitem [{\citenamefont {Lucente}\ \emph {et~al.}(2023)\citenamefont
  {Lucente}, \citenamefont {Viale}, \citenamefont {Gnoli}, \citenamefont
  {Puglisi},\ and\ \citenamefont
  {Vulpiani}}]{lucenteRevealingNonequilibriumNature2023}%
  \BibitemOpen
  \bibfield  {author} {\bibinfo {author} {\bibfnamefont {D.}~\bibnamefont
  {Lucente}}, \bibinfo {author} {\bibfnamefont {M.}~\bibnamefont {Viale}},
  \bibinfo {author} {\bibfnamefont {A.}~\bibnamefont {Gnoli}}, \bibinfo
  {author} {\bibfnamefont {A.}~\bibnamefont {Puglisi}}, \ and\ \bibinfo
  {author} {\bibfnamefont {A.}~\bibnamefont {Vulpiani}},\ }\bibfield  {title}
  {\enquote {\bibinfo {title} {Revealing the {{Nonequilibrium Nature}} of a
  {{Granular Intruder}}: {{The Crucial Role}} of {{Non-Gaussian Behavior}}},}\
  }\href {\doibase 10.1103/PhysRevLett.131.078201} {\bibfield  {journal}
  {\bibinfo  {journal} {Physical Review Letters}\ }\textbf {\bibinfo {volume}
  {131}},\ \bibinfo {pages} {078201} (\bibinfo {year} {2023})}\BibitemShut
  {NoStop}%
\bibitem [{\citenamefont {Gnesotto}\ \emph {et~al.}(2020)\citenamefont
  {Gnesotto}, \citenamefont {Gradziuk}, \citenamefont {Ronceray},\ and\
  \citenamefont {Broedersz}}]{gnesottoLearningNonequilibriumDynamics2020}%
  \BibitemOpen
  \bibfield  {author} {\bibinfo {author} {\bibfnamefont {Federico~S.}\
  \bibnamefont {Gnesotto}}, \bibinfo {author} {\bibfnamefont {Grzegorz}\
  \bibnamefont {Gradziuk}}, \bibinfo {author} {\bibfnamefont {Pierre}\
  \bibnamefont {Ronceray}}, \ and\ \bibinfo {author} {\bibfnamefont {Chase~P.}\
  \bibnamefont {Broedersz}},\ }\bibfield  {title} {\enquote {\bibinfo {title}
  {Learning the non-equilibrium dynamics of {{Brownian}} movies},}\ }\href
  {\doibase 10.1038/s41467-020-18796-9} {\bibfield  {journal} {\bibinfo
  {journal} {Nature Communications}\ }\textbf {\bibinfo {volume} {11}},\
  \bibinfo {pages} {5378} (\bibinfo {year} {2020})}\BibitemShut {NoStop}%
\bibitem [{\citenamefont {Lynn}\ \emph {et~al.}(2022)\citenamefont {Lynn},
  \citenamefont {Holmes}, \citenamefont {Bialek},\ and\ \citenamefont
  {Schwab}}]{lynnDecomposingLocalArrow2022}%
  \BibitemOpen
  \bibfield  {author} {\bibinfo {author} {\bibfnamefont {Christopher~W.}\
  \bibnamefont {Lynn}}, \bibinfo {author} {\bibfnamefont {Caroline~M.}\
  \bibnamefont {Holmes}}, \bibinfo {author} {\bibfnamefont {William}\
  \bibnamefont {Bialek}}, \ and\ \bibinfo {author} {\bibfnamefont {David~J.}\
  \bibnamefont {Schwab}},\ }\bibfield  {title} {\enquote {\bibinfo {title}
  {Decomposing the {{Local Arrow}} of {{Time}} in {{Interacting Systems}}},}\
  }\href {\doibase 10.1103/PhysRevLett.129.118101} {\bibfield  {journal}
  {\bibinfo  {journal} {Physical Review Letters}\ }\textbf {\bibinfo {volume}
  {129}},\ \bibinfo {pages} {118101} (\bibinfo {year} {2022})}\BibitemShut
  {NoStop}%
\bibitem [{\citenamefont {Tan}\ \emph {et~al.}(2021)\citenamefont {Tan},
  \citenamefont {Watson}, \citenamefont {Chao}, \citenamefont {Li},
  \citenamefont {Gingrich}, \citenamefont {Horowitz},\ and\ \citenamefont
  {Fakhri}}]{tanScaledependentIrreversibilityLiving2021}%
  \BibitemOpen
  \bibfield  {author} {\bibinfo {author} {\bibfnamefont {Tzer~Han}\
  \bibnamefont {Tan}}, \bibinfo {author} {\bibfnamefont {Garrett~A.}\
  \bibnamefont {Watson}}, \bibinfo {author} {\bibfnamefont {Yu-Chen}\
  \bibnamefont {Chao}}, \bibinfo {author} {\bibfnamefont {Junang}\ \bibnamefont
  {Li}}, \bibinfo {author} {\bibfnamefont {Todd~R.}\ \bibnamefont {Gingrich}},
  \bibinfo {author} {\bibfnamefont {Jordan~M.}\ \bibnamefont {Horowitz}}, \
  and\ \bibinfo {author} {\bibfnamefont {Nikta}\ \bibnamefont {Fakhri}},\
  }\href@noop {} {\enquote {\bibinfo {title} {Scale-dependent irreversibility
  in living matter},}\ } (\bibinfo {year} {2021}),\ \Eprint
  {http://arxiv.org/abs/2107.05701} {arxiv:2107.05701 [cond-mat,
  physics:physics]} \BibitemShut {NoStop}%
\bibitem [{\citenamefont {Lucente}\ \emph {et~al.}(2022)\citenamefont
  {Lucente}, \citenamefont {Baldassarri}, \citenamefont {Puglisi},
  \citenamefont {Vulpiani},\ and\ \citenamefont
  {Viale}}]{lucenteInferenceTimeIrreversibility2022}%
  \BibitemOpen
  \bibfield  {author} {\bibinfo {author} {\bibfnamefont {D.}~\bibnamefont
  {Lucente}}, \bibinfo {author} {\bibfnamefont {A.}~\bibnamefont
  {Baldassarri}}, \bibinfo {author} {\bibfnamefont {A.}~\bibnamefont
  {Puglisi}}, \bibinfo {author} {\bibfnamefont {A.}~\bibnamefont {Vulpiani}}, \
  and\ \bibinfo {author} {\bibfnamefont {M.}~\bibnamefont {Viale}},\ }\bibfield
   {title} {\enquote {\bibinfo {title} {Inference of time irreversibility from
  incomplete information: {{Linear}} systems and its pitfalls},}\ }\href
  {\doibase 10.1103/PhysRevResearch.4.043103} {\bibfield  {journal} {\bibinfo
  {journal} {Physical Review Research}\ }\textbf {\bibinfo {volume} {4}},\
  \bibinfo {pages} {043103} (\bibinfo {year} {2022})}\BibitemShut {NoStop}%
\bibitem [{\citenamefont {Walldorf}\ \emph {et~al.}(2018)\citenamefont
  {Walldorf}, \citenamefont {Padurariu}, \citenamefont {Jauho},\ and\
  \citenamefont {Flindt}}]{PhysRevLett.120.087701}%
  \BibitemOpen
  \bibfield  {author} {\bibinfo {author} {\bibfnamefont {Nicklas}\ \bibnamefont
  {Walldorf}}, \bibinfo {author} {\bibfnamefont {Ciprian}\ \bibnamefont
  {Padurariu}}, \bibinfo {author} {\bibfnamefont {Antti-Pekka}\ \bibnamefont
  {Jauho}}, \ and\ \bibinfo {author} {\bibfnamefont {Christian}\ \bibnamefont
  {Flindt}},\ }\bibfield  {title} {\enquote {\bibinfo {title} {Electron waiting
  times of a cooper pair splitter},}\ }\href {\doibase
  10.1103/PhysRevLett.120.087701} {\bibfield  {journal} {\bibinfo  {journal}
  {Phys. Rev. Lett.}\ }\textbf {\bibinfo {volume} {120}},\ \bibinfo {pages}
  {087701} (\bibinfo {year} {2018})}\BibitemShut {NoStop}%
\bibitem [{\citenamefont {Landi}(2021)}]{PhysRevB.104.195408}%
  \BibitemOpen
  \bibfield  {author} {\bibinfo {author} {\bibfnamefont {Gabriel~T.}\
  \bibnamefont {Landi}},\ }\bibfield  {title} {\enquote {\bibinfo {title}
  {Waiting time statistics in boundary-driven free fermion chains},}\ }\href
  {\doibase 10.1103/PhysRevB.104.195408} {\bibfield  {journal} {\bibinfo
  {journal} {Phys. Rev. B}\ }\textbf {\bibinfo {volume} {104}},\ \bibinfo
  {pages} {195408} (\bibinfo {year} {2021})}\BibitemShut {NoStop}%
\bibitem [{\citenamefont {Cuetara}\ and\ \citenamefont
  {Esposito}(2015)}]{Cuetara_2015}%
  \BibitemOpen
  \bibfield  {author} {\bibinfo {author} {\bibfnamefont {Gregory~Bulnes}\
  \bibnamefont {Cuetara}}\ and\ \bibinfo {author} {\bibfnamefont
  {Massimiliano}\ \bibnamefont {Esposito}},\ }\bibfield  {title} {\enquote
  {\bibinfo {title} {Double quantum dot coupled to a quantum point contact: a
  stochastic thermodynamics approach},}\ }\href {\doibase
  10.1088/1367-2630/17/9/095005} {\bibfield  {journal} {\bibinfo  {journal}
  {New Journal of Physics}\ }\textbf {\bibinfo {volume} {17}},\ \bibinfo
  {pages} {095005} (\bibinfo {year} {2015})}\BibitemShut {NoStop}%
\bibitem [{\citenamefont {Viisanen}\ \emph {et~al.}(2015)\citenamefont
  {Viisanen}, \citenamefont {Suomela}, \citenamefont {Gasparinetti},
  \citenamefont {Saira}, \citenamefont {Ankerhold},\ and\ \citenamefont
  {Pekola}}]{Viisanen_2015}%
  \BibitemOpen
  \bibfield  {author} {\bibinfo {author} {\bibfnamefont {Klaara~L}\
  \bibnamefont {Viisanen}}, \bibinfo {author} {\bibfnamefont {Samu}\
  \bibnamefont {Suomela}}, \bibinfo {author} {\bibfnamefont {Simone}\
  \bibnamefont {Gasparinetti}}, \bibinfo {author} {\bibfnamefont {Olli-Pentti}\
  \bibnamefont {Saira}}, \bibinfo {author} {\bibfnamefont {Joachim}\
  \bibnamefont {Ankerhold}}, \ and\ \bibinfo {author} {\bibfnamefont {Jukka~P}\
  \bibnamefont {Pekola}},\ }\bibfield  {title} {\enquote {\bibinfo {title}
  {Incomplete measurement of work in a dissipative two level system},}\ }\href
  {\doibase 10.1088/1367-2630/17/5/055014} {\bibfield  {journal} {\bibinfo
  {journal} {New Journal of Physics}\ }\textbf {\bibinfo {volume} {17}},\
  \bibinfo {pages} {055014} (\bibinfo {year} {2015})}\BibitemShut {NoStop}%
\bibitem [{\citenamefont {S\'anchez}\ \emph {et~al.}(2019)\citenamefont
  {S\'anchez}, \citenamefont {Splettstoesser},\ and\ \citenamefont
  {Whitney}}]{PhysRevLett.123.216801}%
  \BibitemOpen
  \bibfield  {author} {\bibinfo {author} {\bibfnamefont {Rafael}\ \bibnamefont
  {S\'anchez}}, \bibinfo {author} {\bibfnamefont {Janine}\ \bibnamefont
  {Splettstoesser}}, \ and\ \bibinfo {author} {\bibfnamefont {Robert~S.}\
  \bibnamefont {Whitney}},\ }\bibfield  {title} {\enquote {\bibinfo {title}
  {Nonequilibrium system as a demon},}\ }\href {\doibase
  10.1103/PhysRevLett.123.216801} {\bibfield  {journal} {\bibinfo  {journal}
  {Phys. Rev. Lett.}\ }\textbf {\bibinfo {volume} {123}},\ \bibinfo {pages}
  {216801} (\bibinfo {year} {2019})}\BibitemShut {NoStop}%
\bibitem [{\citenamefont {Freitas}\ and\ \citenamefont
  {Esposito}(2021)}]{PhysRevE.103.032118}%
  \BibitemOpen
  \bibfield  {author} {\bibinfo {author} {\bibfnamefont {Nahuel}\ \bibnamefont
  {Freitas}}\ and\ \bibinfo {author} {\bibfnamefont {Massimiliano}\
  \bibnamefont {Esposito}},\ }\bibfield  {title} {\enquote {\bibinfo {title}
  {Characterizing autonomous maxwell demons},}\ }\href {\doibase
  10.1103/PhysRevE.103.032118} {\bibfield  {journal} {\bibinfo  {journal}
  {Phys. Rev. E}\ }\textbf {\bibinfo {volume} {103}},\ \bibinfo {pages}
  {032118} (\bibinfo {year} {2021})}\BibitemShut {NoStop}%
\bibitem [{\citenamefont {Rao}\ and\ \citenamefont
  {Esposito}(2018{\natexlab{a}})}]{rao2018conservation}%
  \BibitemOpen
  \bibfield  {author} {\bibinfo {author} {\bibfnamefont {Riccardo}\
  \bibnamefont {Rao}}\ and\ \bibinfo {author} {\bibfnamefont {Massimiliano}\
  \bibnamefont {Esposito}},\ }\bibfield  {title} {\enquote {\bibinfo {title}
  {Conservation laws and work fluctuation relations in chemical reaction
  networks},}\ }\href {\doibase https://doi.org/10.1063/1.5042253} {\bibfield
  {journal} {\bibinfo  {journal} {The Journal of chemical physics}\ }\textbf
  {\bibinfo {volume} {149}} (\bibinfo {year} {2018}{\natexlab{a}}),\
  https://doi.org/10.1063/1.5042253}\BibitemShut {NoStop}%
\bibitem [{\citenamefont {Bierbaum}\ and\ \citenamefont
  {Lipowsky}(2011)}]{bierbaumChemomechanicalCouplingMotor2011}%
  \BibitemOpen
  \bibfield  {author} {\bibinfo {author} {\bibfnamefont {Veronika}\
  \bibnamefont {Bierbaum}}\ and\ \bibinfo {author} {\bibfnamefont {Reinhard}\
  \bibnamefont {Lipowsky}},\ }\bibfield  {title} {\enquote {\bibinfo {title}
  {Chemomechanical {{Coupling}} and {{Motor Cycles}} of {{Myosin V}}},}\ }\href
  {\doibase 10.1016/j.bpj.2011.02.012} {\bibfield  {journal} {\bibinfo
  {journal} {Biophysical Journal}\ }\textbf {\bibinfo {volume} {100}},\
  \bibinfo {pages} {1747--1755} (\bibinfo {year} {2011})}\BibitemShut {NoStop}%
\bibitem [{\citenamefont {Abbondanzieri}\ \emph {et~al.}(2005)\citenamefont
  {Abbondanzieri}, \citenamefont {Greenleaf}, \citenamefont {Shaevitz},
  \citenamefont {Landick},\ and\ \citenamefont
  {Block}}]{abbondanzieriDirectObservationBasepair2005}%
  \BibitemOpen
  \bibfield  {author} {\bibinfo {author} {\bibfnamefont {Elio~A.}\ \bibnamefont
  {Abbondanzieri}}, \bibinfo {author} {\bibfnamefont {William~J.}\ \bibnamefont
  {Greenleaf}}, \bibinfo {author} {\bibfnamefont {Joshua~W.}\ \bibnamefont
  {Shaevitz}}, \bibinfo {author} {\bibfnamefont {Robert}\ \bibnamefont
  {Landick}}, \ and\ \bibinfo {author} {\bibfnamefont {Steven~M.}\ \bibnamefont
  {Block}},\ }\bibfield  {title} {\enquote {\bibinfo {title} {Direct
  observation of base-pair stepping by {{RNA}} polymerase},}\ }\href {\doibase
  10.1038/nature04268} {\bibfield  {journal} {\bibinfo  {journal} {Nature}\
  }\textbf {\bibinfo {volume} {438}},\ \bibinfo {pages} {460--465} (\bibinfo
  {year} {2005})}\BibitemShut {NoStop}%
\bibitem [{\citenamefont {Chemla}\ \emph {et~al.}(2008)\citenamefont {Chemla},
  \citenamefont {Moffitt},\ and\ \citenamefont
  {Bustamante}}]{chemlaExactSolutionsKinetic2008}%
  \BibitemOpen
  \bibfield  {author} {\bibinfo {author} {\bibfnamefont {Yann~R.}\ \bibnamefont
  {Chemla}}, \bibinfo {author} {\bibfnamefont {Jeffrey~R.}\ \bibnamefont
  {Moffitt}}, \ and\ \bibinfo {author} {\bibfnamefont {Carlos}\ \bibnamefont
  {Bustamante}},\ }\bibfield  {title} {\enquote {\bibinfo {title} {Exact
  {{Solutions}} for {{Kinetic Models}} of {{Macromolecular Dynamics}}},}\
  }\href {\doibase 10.1021/jp076153r} {\bibfield  {journal} {\bibinfo
  {journal} {The Journal of Physical Chemistry B}\ }\textbf {\bibinfo {volume}
  {112}},\ \bibinfo {pages} {6025--6044} (\bibinfo {year} {2008})}\BibitemShut
  {NoStop}%
\bibitem [{Note1()}]{Note1}%
  \BibitemOpen
  \bibinfo {note} {Connecting to the notation of Ref.~\cite
  {PhysRevX.12.041026}, \(\protect \bm {\Delta }_\ell \) can be expressed as \(
  \vert \protect \hspace {-.45ex} \vert \ell \rangle \protect \hspace {-.7ex}
  \rangle \langle \protect \hspace {-.7ex} \langle \ell \vert \protect \hspace
  {-.45ex} \vert \).}\BibitemShut {Stop}%
\bibitem [{\citenamefont {Sekimoto}(2021)}]{sekimoto2021derivation}%
  \BibitemOpen
  \bibfield  {author} {\bibinfo {author} {\bibfnamefont {Ken}\ \bibnamefont
  {Sekimoto}},\ }\bibfield  {title} {\enquote {\bibinfo {title} {Derivation of
  the first passage time distribution for markovian process on discrete
  network},}\ }\href {https://arxiv.org/abs/2110.02216} {\  (\bibinfo {year}
  {2021})},\ \Eprint {http://arxiv.org/abs/2110.02216} {arXiv:2110.02216
  [cond-mat.stat-mech]} \BibitemShut {NoStop}%
\bibitem [{\citenamefont {Schnakenberg}(1976)}]{RevModPhys.48.571}%
  \BibitemOpen
  \bibfield  {author} {\bibinfo {author} {\bibfnamefont {J.}~\bibnamefont
  {Schnakenberg}},\ }\bibfield  {title} {\enquote {\bibinfo {title} {Network
  theory of microscopic and macroscopic behavior of master equation systems},}\
  }\href {\doibase 10.1103/RevModPhys.48.571} {\bibfield  {journal} {\bibinfo
  {journal} {Rev. Mod. Phys.}\ }\textbf {\bibinfo {volume} {48}},\ \bibinfo
  {pages} {571--585} (\bibinfo {year} {1976})}\BibitemShut {NoStop}%
\bibitem [{Note2()}]{Note2}%
  \BibitemOpen
  \bibinfo {note} {Empirical estimation of the divergence of continuous
  variables presents convergence problems, the reader may refer to the code
  in~\cite {Harunari_KLD_estimation} for an unbiased estimator.}\BibitemShut
  {Stop}%
\bibitem [{\citenamefont {Falasco}\ and\ \citenamefont
  {Esposito}(2021)}]{falascoLocalDetailedBalance2021a}%
  \BibitemOpen
  \bibfield  {author} {\bibinfo {author} {\bibfnamefont {Gianmaria}\
  \bibnamefont {Falasco}}\ and\ \bibinfo {author} {\bibfnamefont
  {Massimiliano}\ \bibnamefont {Esposito}},\ }\bibfield  {title} {\enquote
  {\bibinfo {title} {Local detailed balance across scales: {{From}} diffusions
  to jump processes and beyond},}\ }\href {\doibase
  10.1103/PhysRevE.103.042114} {\bibfield  {journal} {\bibinfo  {journal}
  {Physical Review E}\ }\textbf {\bibinfo {volume} {103}},\ \bibinfo {pages}
  {042114} (\bibinfo {year} {2021})}\BibitemShut {NoStop}%
\bibitem [{Note3()}]{Note3}%
  \BibitemOpen
  \bibinfo {note} {Local detailed balance for a given transition reads \(
  r_\ell / r_{\protect \overline {\ell }} = \exp (-\phi _\ell )\), with \(\phi
  _\ell \) the entropy flux generated by \(\ell \), and the affinity of a cycle
  becomes \(\DOTSB \sum@ \slimits@ _{\ell \in \protect \vec {\protect \mathcal
  {C}}} \phi _\ell \).}\BibitemShut {Stop}%
\bibitem [{\citenamefont {Polettini}\ \emph {et~al.}(2016)\citenamefont
  {Polettini}, \citenamefont {{Bulnes-Cuetara}},\ and\ \citenamefont
  {Esposito}}]{polettiniConservationLawsSymmetries2016}%
  \BibitemOpen
  \bibfield  {author} {\bibinfo {author} {\bibfnamefont {Matteo}\ \bibnamefont
  {Polettini}}, \bibinfo {author} {\bibfnamefont {Gregory}\ \bibnamefont
  {{Bulnes-Cuetara}}}, \ and\ \bibinfo {author} {\bibfnamefont {Massimiliano}\
  \bibnamefont {Esposito}},\ }\bibfield  {title} {\enquote {\bibinfo {title}
  {Conservation laws and symmetries in stochastic thermodynamics},}\ }\href
  {\doibase 10.1103/PhysRevE.94.052117} {\bibfield  {journal} {\bibinfo
  {journal} {Physical Review E}\ }\textbf {\bibinfo {volume} {94}},\ \bibinfo
  {pages} {052117} (\bibinfo {year} {2016})}\BibitemShut {NoStop}%
\bibitem [{\citenamefont {Rao}\ and\ \citenamefont
  {Esposito}(2018{\natexlab{b}})}]{Rao_2018}%
  \BibitemOpen
  \bibfield  {author} {\bibinfo {author} {\bibfnamefont {Riccardo}\
  \bibnamefont {Rao}}\ and\ \bibinfo {author} {\bibfnamefont {Massimiliano}\
  \bibnamefont {Esposito}},\ }\bibfield  {title} {\enquote {\bibinfo {title}
  {Conservation laws shape dissipation},}\ }\href {\doibase
  10.1088/1367-2630/aaa15f} {\bibfield  {journal} {\bibinfo  {journal} {New
  Journal of Physics}\ }\textbf {\bibinfo {volume} {20}},\ \bibinfo {pages}
  {023007} (\bibinfo {year} {2018}{\natexlab{b}})}\BibitemShut {NoStop}%
\bibitem [{\citenamefont {Falasco}\ and\ \citenamefont
  {Esposito}(2023)}]{falascoMacroscopicStochasticThermodynamics2023}%
  \BibitemOpen
  \bibfield  {author} {\bibinfo {author} {\bibfnamefont {Gianmaria}\
  \bibnamefont {Falasco}}\ and\ \bibinfo {author} {\bibfnamefont
  {Massimiliano}\ \bibnamefont {Esposito}},\ }\href {\doibase
  10.48550/arXiv.2307.12406} {\enquote {\bibinfo {title} {Macroscopic
  {{Stochastic Thermodynamics}}},}\ } (\bibinfo {year} {2023}),\ \Eprint
  {http://arxiv.org/abs/2307.12406} {arxiv:2307.12406 [cond-mat,
  physics:math-ph]} \BibitemShut {NoStop}%
\bibitem [{\citenamefont {Avanzini}\ \emph {et~al.}(2024)\citenamefont
  {Avanzini}, \citenamefont {Bilancioni}, \citenamefont {Cavina}, \citenamefont
  {Dal~Cengio}, \citenamefont {Esposito}, \citenamefont {Falasco},
  \citenamefont {Forastiere}, \citenamefont {Freitas}, \citenamefont {Garilli},
  \citenamefont {Harunari}, \citenamefont {Lecomte}, \citenamefont {Lazarescu},
  \citenamefont {Marehalli~Srinivas}, \citenamefont {Moslonka}, \citenamefont
  {Neri}, \citenamefont {Penocchio}, \citenamefont {Pi{\~n}eros}, \citenamefont
  {Polettini}, \citenamefont {Raghu}, \citenamefont {Raux}, \citenamefont
  {Sekimoto},\ and\ \citenamefont {Soret}}]{avanzini_methods_2023}%
  \BibitemOpen
  \bibfield  {author} {\bibinfo {author} {\bibfnamefont {Francesco}\
  \bibnamefont {Avanzini}}, \bibinfo {author} {\bibfnamefont {Massimo}\
  \bibnamefont {Bilancioni}}, \bibinfo {author} {\bibfnamefont {Vasco}\
  \bibnamefont {Cavina}}, \bibinfo {author} {\bibfnamefont {Sara}\ \bibnamefont
  {Dal~Cengio}}, \bibinfo {author} {\bibfnamefont {Massimiliano}\ \bibnamefont
  {Esposito}}, \bibinfo {author} {\bibfnamefont {Gianmaria}\ \bibnamefont
  {Falasco}}, \bibinfo {author} {\bibfnamefont {Danilo}\ \bibnamefont
  {Forastiere}}, \bibinfo {author} {\bibfnamefont {Jose~Nahuel}\ \bibnamefont
  {Freitas}}, \bibinfo {author} {\bibfnamefont {Alberto}\ \bibnamefont
  {Garilli}}, \bibinfo {author} {\bibfnamefont {Pedro~E.}\ \bibnamefont
  {Harunari}}, \bibinfo {author} {\bibfnamefont {Vivien}\ \bibnamefont
  {Lecomte}}, \bibinfo {author} {\bibfnamefont {Alexandre}\ \bibnamefont
  {Lazarescu}}, \bibinfo {author} {\bibfnamefont {Shesha~G.}\ \bibnamefont
  {Marehalli~Srinivas}}, \bibinfo {author} {\bibfnamefont {Charles}\
  \bibnamefont {Moslonka}}, \bibinfo {author} {\bibfnamefont {Izaak}\
  \bibnamefont {Neri}}, \bibinfo {author} {\bibfnamefont {Emanuele}\
  \bibnamefont {Penocchio}}, \bibinfo {author} {\bibfnamefont {William~D.}\
  \bibnamefont {Pi{\~n}eros}}, \bibinfo {author} {\bibfnamefont {Matteo}\
  \bibnamefont {Polettini}}, \bibinfo {author} {\bibfnamefont {Adarsh}\
  \bibnamefont {Raghu}}, \bibinfo {author} {\bibfnamefont {Paul}\ \bibnamefont
  {Raux}}, \bibinfo {author} {\bibfnamefont {Ken}\ \bibnamefont {Sekimoto}}, \
  and\ \bibinfo {author} {\bibfnamefont {Ariane}\ \bibnamefont {Soret}},\
  }\bibfield  {title} {\enquote {\bibinfo {title} {Methods and conversations in
  (post)modern thermodynamics},}\ }\href {\doibase
  10.21468/SciPostPhysLectNotes.80} {\bibfield  {journal} {\bibinfo  {journal}
  {SciPost Physics Lecture Notes}\ ,\ \bibinfo {pages} {080}} (\bibinfo {year}
  {2024})}\BibitemShut {NoStop}%
\bibitem [{\citenamefont {Schl{\"o}gl}(1972)}]{10__Schlogl1972}%
  \BibitemOpen
  \bibfield  {author} {\bibinfo {author} {\bibfnamefont {F.}~\bibnamefont
  {Schl{\"o}gl}},\ }\bibfield  {title} {\enquote {\bibinfo {title} {Chemical
  reaction models for non-equilibrium phase transitions},}\ }\href {\doibase
  10.1007/BF01379769} {\bibfield  {journal} {\bibinfo  {journal} {Zeitschrift
  f{\"u}r Physik}\ }\textbf {\bibinfo {volume} {253}},\ \bibinfo {pages}
  {147--161} (\bibinfo {year} {1972})}\BibitemShut {NoStop}%
\bibitem [{\citenamefont {Freitas}\ \emph {et~al.}(2021)\citenamefont
  {Freitas}, \citenamefont {Delvenne},\ and\ \citenamefont
  {Esposito}}]{freitasStochasticThermodynamicsNonlinear2021}%
  \BibitemOpen
  \bibfield  {author} {\bibinfo {author} {\bibfnamefont {Nahuel}\ \bibnamefont
  {Freitas}}, \bibinfo {author} {\bibfnamefont {Jean-Charles}\ \bibnamefont
  {Delvenne}}, \ and\ \bibinfo {author} {\bibfnamefont {Massimiliano}\
  \bibnamefont {Esposito}},\ }\bibfield  {title} {\enquote {\bibinfo {title}
  {Stochastic {{Thermodynamics}} of {{Nonlinear Electronic Circuits}}: {{A
  Realistic Framework}} for {{Computing Around}} \${{kT}}\$},}\ }\href
  {\doibase 10.1103/PhysRevX.11.031064} {\bibfield  {journal} {\bibinfo
  {journal} {Physical Review X}\ }\textbf {\bibinfo {volume} {11}},\ \bibinfo
  {pages} {031064} (\bibinfo {year} {2021})}\BibitemShut {NoStop}%
\bibitem [{\citenamefont {Vroylandt}\ \emph {et~al.}(2017)\citenamefont
  {Vroylandt}, \citenamefont {Esposito},\ and\ \citenamefont
  {Verley}}]{vroylandtCollectiveEffectsEnhancing2017}%
  \BibitemOpen
  \bibfield  {author} {\bibinfo {author} {\bibfnamefont {Hadrien}\ \bibnamefont
  {Vroylandt}}, \bibinfo {author} {\bibfnamefont {Massimiliano}\ \bibnamefont
  {Esposito}}, \ and\ \bibinfo {author} {\bibfnamefont {Gatien}\ \bibnamefont
  {Verley}},\ }\bibfield  {title} {\enquote {\bibinfo {title} {Collective
  effects enhancing power and efficiency},}\ }\href {\doibase
  10.1209/0295-5075/120/30009} {\bibfield  {journal} {\bibinfo  {journal} {EPL
  (Europhysics Letters)}\ }\textbf {\bibinfo {volume} {120}},\ \bibinfo {pages}
  {30009} (\bibinfo {year} {2017})}\BibitemShut {NoStop}%
\bibitem [{\citenamefont {Borrelli}\ \emph {et~al.}(2015)\citenamefont
  {Borrelli}, \citenamefont {Koski}, \citenamefont {Maniscalco},\ and\
  \citenamefont {Pekola}}]{PhysRevE.91.012145}%
  \BibitemOpen
  \bibfield  {author} {\bibinfo {author} {\bibfnamefont {Massimo}\ \bibnamefont
  {Borrelli}}, \bibinfo {author} {\bibfnamefont {Jonne~V.}\ \bibnamefont
  {Koski}}, \bibinfo {author} {\bibfnamefont {Sabrina}\ \bibnamefont
  {Maniscalco}}, \ and\ \bibinfo {author} {\bibfnamefont {Jukka~P.}\
  \bibnamefont {Pekola}},\ }\bibfield  {title} {\enquote {\bibinfo {title}
  {Fluctuation relations for driven coupled classical two-level systems with
  incomplete measurements},}\ }\href {\doibase 10.1103/PhysRevE.91.012145}
  {\bibfield  {journal} {\bibinfo  {journal} {Phys. Rev. E}\ }\textbf {\bibinfo
  {volume} {91}},\ \bibinfo {pages} {012145} (\bibinfo {year}
  {2015})}\BibitemShut {NoStop}%
\bibitem [{\citenamefont {Fritz}\ \emph {et~al.}(2020)\citenamefont {Fritz},
  \citenamefont {Nguyen},\ and\ \citenamefont
  {Seifert}}]{fritz_stochastic_2020}%
  \BibitemOpen
  \bibfield  {author} {\bibinfo {author} {\bibfnamefont {Jonas~H.}\
  \bibnamefont {Fritz}}, \bibinfo {author} {\bibfnamefont {Basile}\
  \bibnamefont {Nguyen}}, \ and\ \bibinfo {author} {\bibfnamefont {Udo}\
  \bibnamefont {Seifert}},\ }\bibfield  {title} {\enquote {\bibinfo {title}
  {Stochastic thermodynamics of chemical reactions coupled to finite
  reservoirs: {A} case study for the {Brusselator}},}\ }\href {\doibase
  10.1063/5.0006115} {\bibfield  {journal} {\bibinfo  {journal} {J. Chem.
  Phys.}\ }\textbf {\bibinfo {volume} {152}},\ \bibinfo {pages} {235101}
  (\bibinfo {year} {2020})}\BibitemShut {NoStop}%
\bibitem [{\citenamefont {Nicolis}\ and\ \citenamefont
  {Prigogine}(1977)}]{nicolis_self-organization_1977}%
  \BibitemOpen
  \bibfield  {author} {\bibinfo {author} {\bibfnamefont {G.}~\bibnamefont
  {Nicolis}}\ and\ \bibinfo {author} {\bibfnamefont {I.}~\bibnamefont
  {Prigogine}},\ }\href@noop {} {\emph {\bibinfo {title} {Self-{Organization}
  in {Nonequilibrium} {Systems}: {From} {Dissipative} {Structures} to {Order}
  {Through} {Fluctuations}}}},\ A {Wiley}-{Interscience} publication\ (\bibinfo
   {publisher} {Wiley},\ \bibinfo {year} {1977})\BibitemShut {NoStop}%
\bibitem [{\citenamefont {Qian}\ \emph {et~al.}(2002)\citenamefont {Qian},
  \citenamefont {Saffarian},\ and\ \citenamefont
  {Elson}}]{qian2002concentration}%
  \BibitemOpen
  \bibfield  {author} {\bibinfo {author} {\bibfnamefont {Hong}\ \bibnamefont
  {Qian}}, \bibinfo {author} {\bibfnamefont {Saveez}\ \bibnamefont
  {Saffarian}}, \ and\ \bibinfo {author} {\bibfnamefont {Elliot~L}\
  \bibnamefont {Elson}},\ }\bibfield  {title} {\enquote {\bibinfo {title}
  {Concentration fluctuations in a mesoscopic oscillating chemical reaction
  system},}\ }\href {\doibase https://doi.org/10.1073/pnas.152007599}
  {\bibfield  {journal} {\bibinfo  {journal} {Proceedings of the National
  Academy of Sciences}\ }\textbf {\bibinfo {volume} {99}},\ \bibinfo {pages}
  {10376--10381} (\bibinfo {year} {2002})}\BibitemShut {NoStop}%
\bibitem [{\citenamefont {Nguyen}\ \emph {et~al.}(2018)\citenamefont {Nguyen},
  \citenamefont {Seifert},\ and\ \citenamefont {Barato}}]{nguyen2018phase}%
  \BibitemOpen
  \bibfield  {author} {\bibinfo {author} {\bibfnamefont {Basile}\ \bibnamefont
  {Nguyen}}, \bibinfo {author} {\bibfnamefont {Udo}\ \bibnamefont {Seifert}}, \
  and\ \bibinfo {author} {\bibfnamefont {Andre~C}\ \bibnamefont {Barato}},\
  }\bibfield  {title} {\enquote {\bibinfo {title} {Phase transition in
  thermodynamically consistent biochemical oscillators},}\ }\href
  {https://doi.org/10.1063/1.5032104} {\bibfield  {journal} {\bibinfo
  {journal} {The Journal of Chemical Physics}\ }\textbf {\bibinfo {volume}
  {149}} (\bibinfo {year} {2018})}\BibitemShut {NoStop}%
\bibitem [{\citenamefont {Fernandes~Martins}\ and\ \citenamefont
  {Horowitz}(2023)}]{fernandesmartinsTopologicallyConstrainedFluctuations2023}%
  \BibitemOpen
  \bibfield  {author} {\bibinfo {author} {\bibfnamefont {Gabriela}\
  \bibnamefont {Fernandes~Martins}}\ and\ \bibinfo {author} {\bibfnamefont
  {Jordan~M.}\ \bibnamefont {Horowitz}},\ }\bibfield  {title} {\enquote
  {\bibinfo {title} {Topologically constrained fluctuations and thermodynamics
  regulate nonequilibrium response},}\ }\href {\doibase
  10.1103/PhysRevE.108.044113} {\bibfield  {journal} {\bibinfo  {journal}
  {Physical Review E}\ }\textbf {\bibinfo {volume} {108}},\ \bibinfo {pages}
  {044113} (\bibinfo {year} {2023})}\BibitemShut {NoStop}%
\bibitem [{\citenamefont
  {Sch{\"o}nenberger}(2020)}]{schonenbergerFoldingEnergyKinetics}%
  \BibitemOpen
  \bibfield  {author} {\bibinfo {author} {\bibfnamefont {David}\ \bibnamefont
  {Sch{\"o}nenberger}},\ }\bibfield  {title} {\enquote {\bibinfo {title}
  {Folding energy and kinetics of mutually interacting {{DNA Hairpins}}},}\
  }\href {http://hdl.handle.net/2445/171468} {\  (\bibinfo {year}
  {2020})}\BibitemShut {NoStop}%
\bibitem [{\citenamefont {Dal~Cengio}\ \emph {et~al.}(2023)\citenamefont
  {Dal~Cengio}, \citenamefont {Lecomte},\ and\ \citenamefont
  {Polettini}}]{Cengio2022}%
  \BibitemOpen
  \bibfield  {author} {\bibinfo {author} {\bibfnamefont {Sara}\ \bibnamefont
  {Dal~Cengio}}, \bibinfo {author} {\bibfnamefont {Vivien}\ \bibnamefont
  {Lecomte}}, \ and\ \bibinfo {author} {\bibfnamefont {Matteo}\ \bibnamefont
  {Polettini}},\ }\bibfield  {title} {\enquote {\bibinfo {title} {Geometry of
  nonequilibrium reaction networks},}\ }\href {\doibase
  10.1103/PhysRevX.13.021040} {\bibfield  {journal} {\bibinfo  {journal}
  {Physical Review X}\ }\textbf {\bibinfo {volume} {13}},\ \bibinfo {pages}
  {021040} (\bibinfo {year} {2023})}\BibitemShut {NoStop}%
\bibitem [{\citenamefont {Harunari}(2024)}]{github_multifilar}%
  \BibitemOpen
  \bibfield  {author} {\bibinfo {author} {\bibfnamefont {Pedro~E.}\
  \bibnamefont {Harunari}},\ }\href {github.com/pedroharunari/MultifilarEvents}
  {\enquote {\bibinfo {title} {Multifilarevents},}\ }\bibinfo {howpublished}
  {\url{https://github.com/pedroharunari/MultifilarEvents}} (\bibinfo {year}
  {2024})\BibitemShut {NoStop}%
\bibitem [{\citenamefont {Gershgorin}(1931)}]{gershgorin1931uber}%
  \BibitemOpen
  \bibfield  {author} {\bibinfo {author} {\bibfnamefont {S.}~\bibnamefont
  {Gershgorin}},\ }\bibfield  {title} {\enquote {\bibinfo {title} {\"uber die
  abgrenzung der eigenwerte einer matrix},}\ }\href
  {https://www.mathnet.ru/rus/im5235} {\bibfield  {journal} {\bibinfo
  {journal} {Bull. Acad. Sci. URSS}\ ,\ \bibinfo {pages} {749--754}} (\bibinfo
  {year} {1931})}\BibitemShut {NoStop}%
\bibitem [{\citenamefont {Harunari}\ and\ \citenamefont
  {Yssou}(2022)}]{Harunari_KLD_estimation}%
  \BibitemOpen
  \bibfield  {author} {\bibinfo {author} {\bibfnamefont {Pedro~E.}\
  \bibnamefont {Harunari}}\ and\ \bibinfo {author} {\bibfnamefont {Ariel}\
  \bibnamefont {Yssou}},\ }\href
  {https://github.com/pedroharunari/KLD_estimation} {\enquote {\bibinfo {title}
  {Kullback-leibler divergence estimation algorithm and inter-transition times
  application},}\ }\bibinfo {howpublished}
  {\url{https://github.com/pedroharunari/KLD_estimation}} (\bibinfo {year}
  {2022})\BibitemShut {NoStop}%
\end{thebibliography}%

\end{document}